\documentstyle[12pt]{article}

\newtheorem{theorem}{Theorem}
\newtheorem{corollary}[theorem]{Corollary}
\newtheorem{lemma}[theorem]{Lemma}

\newcommand{\be}{\begin{equation}}
\newcommand{\ee}{\end{equation}}
\newcommand{\bea}{\begin{eqnarray}}
\newcommand{\eea}{\end{eqnarray}}
\newcommand{\barr}{\begin{array}}
\newcommand{\earr}{\end{array}}

\newcommand{\benum}{\begin{enumerate}}
\newcommand{\eenum}{\end{enumerate}}

\newcommand{\blem}{\begin{lemma}}
\newcommand{\elem}{\end{lemma}}
\newcommand{\bthm}{\begin{theorem}}
\newcommand{\ethm}{\end{theorem}}
\newcommand{\bcor}{\begin{corollary}}
\newcommand{\ecor}{\end{corollary}}

\newcommand{\nn}{\nonumber}
\newcommand{\nind}{\noindent}
\newcommand{\Proof}{\nind{\bf Proof:}\quad}
\newcommand{\pr}{\nind{\bf Proof:}\quad}
\newcommand{\re}{\nind{\bf Remark:}\quad}

\newcommand{\QED}{$\hfill\Box$}
\newcommand{\pa}{\partial}
\newcommand{\one}{\cO(1)}

\newcommand{\al}{\alpha}
\newcommand{\De}{\Delta}
\newcommand{\de}{\delta}
\newcommand{\g}{\gamma}
\newcommand{\G}{\Gamma}
\newcommand{\f}{\phi}

\newcommand{\z}{\zeta}
\newcommand{\r}{\rho}
\newcommand{\k}{\kappa}
\newcommand{\La}{\Lambda}
\newcommand{\la}{\lambda}

\newcommand{\Th}{\Theta}
\newcommand{\th}{\theta}
\newcommand{\ep}{\epsilon}
\newcommand{\si}{\sigma}

\newcommand{\cC}{{\cal C}}
\newcommand{\cZ}{{\cal Z}}
\newcommand{\cO}{{\cal O}}
\newcommand{\cS}{{\cal S}}
\newcommand{\cE}{{\cal E}}
\newcommand{\cR}{{\cal R}}
\newcommand{\cF}{{\cal F}}

\newcommand{\bR}{{\bf R}}
\newcommand{\bZ}{{\bf Z}}

\newcommand{\imag}{{\rm Im}}

\begin{document}
\nonstopmode
\bibliographystyle{plain}

\title{Sine-Gordon   revisited}
\author{J. Dimock\thanks{Research supported by NSF Grant PHY9722045}\\
Dept. of Mathematics \\
SUNY at Buffalo \\
Buffalo, N.Y. 14214 \and T. R. Hurd\thanks{Research supported by the
Natural Sciences and Engineering Research Council of Canada.}\\Dept. of
Mathematics and Statistics\\McMaster University\\Hamilton,
Ontario, L8S 4K1}
\maketitle

\begin{abstract}
We study the
sine-Gordon model in two dimensional space time in two
different domains.  For
$\beta  > 8 \pi $ and weak coupling, we introduce an ultraviolet cutoff
and study the infrared behavior.  A renormalization group analysis
shows that the the model is asymptotically  free in the infrared.
 For $\beta  < 8 \pi $ and weak coupling, we introduce an infrared cutoff
and study the ultraviolet behavior.  A renormalization group analysis
shows that the model is asymptotically  free in the ultraviolet.

\end{abstract}
\newpage

\tableofcontents

\newpage

\section{Introduction}

We are concerned with the two dimensional sine-Gordon model.  The model is
characterized by its partition function which is formally
\be Z = \int  \exp  \left(
   \z \int  \cos (\phi(x) ) dx  \ - \ \frac {1}{2\beta}  \int |\pa
\phi(x)|^2 dx
\right)
\prod_{x \in \bR^2}  d \phi(x)  \ee
It is of interest both as a Euclidean  quantum field theory and because it
describes the classical statistical mechanics of a Coulomb
gas with inverse temperature $\beta$ and activity $\z/2$.

The expression for $Z$ is ill-defined. To  make sense of it
we first replace the plane $\bR^2$ by the torus  $\La_M  =   \bR^2 / L^{M}
\bZ^2$,
where $M$ is a non-negative integer and $L$ is a fixed large positive
constant.
Then the
quadratic term is combined with the non-existent  Lebesgue measure to
  give
a  Gaussian measure.  Introduction of a short distance
cutoff at scale $L^{-N}$ gives the Gaussian measure  $\mu_{\beta v^M_{-N}}$
with  covariance
\begin{equation}  \label{vdef}
\beta v^M_{-N}( x-y)=\beta |\La_M|^{-1}\sum_{p\in \La_M^*, p \neq 0}
\frac{e^{ip(x-y)}}{p^2}e^{-p^4L^{-4N}}
\end{equation}
where  $\La_M^*  = ( 2 \pi L)^{-M} \bZ^2 $.
Since $p=0$ is excluded the measure is supported on fields $\phi$
with  $\int \phi =0$.  Furthermore the  covariance
is smooth and so the measure
is supported  on smooth functions.
    Thus the cutoff  expression
\begin{equation}  \label{Z0}
Z=\int \exp\left(\z\int_{\La_M}\ \cos(\f(x))dx\right)\ d\mu_{\beta
v^M_{-N}}(\f).
\end{equation}
is well-defined.
We are interested in studying the limits $N \to \infty $ (the UV
problem) and $M \to \infty$ (the IR problem).

There are two distinct domains in which these problems are
tractable.   For $\beta > 8 \pi$ and $\zeta$ small it turns out
that the long distance
behavior differs only slightly from that of the free model ($\zeta =0$)
and thus the IR problem can be  controlled.   For $\beta < 8 \pi$
and $\zeta$ small it turns out that
the short distance behavior differs only slightly from free and
thus the UV problem  can be controlled. The purpose of this paper is to
carry  out the analysis  in each case using a renormalization group  (RG)
method.

Each of these problems have been previously studied by the authors
in  \cite{DiHu91} and \cite{DiHu93}.   Unfortunately
there is an error which  occurs in both papers and spoils the proofs of the
main
results.\footnote{The problem is that for the homotopy property one needs
$\k$ small, but the limitation on $\k$ cannot be made independently
of $L$ as was implicitly assumed.
In fact one needs $\kappa \leq
\cO(L^{-2})$  or smaller.  Then the use of  Sobolev
inequalities require $\kappa ( h_1^*)^2 \geq \one $ and hence $h_1^* \geq
\cO(L)$.
This spoils the estimate above line (49) in \cite{DiHu93}.
There is a similar problem in  \cite{DiHu91}.  }
 In our present paper we are at last able to fix this error, and
reinstate our earlier
results. The fix requires some substantial modifications to the method, and
so we give here
reasonably self-contained proofs of the main technical lemmas.
\bigskip

We first discuss the IR problem for $\beta > 8 \pi$.
  We study  the expression (\ref{Z0})
with the ultraviolet cutoff $N$ fixed: for simplicity we take
$N=0$.   The RG method involves
 the introduction of a sum over
scales.  For any   $0 \leq j \leq M$ we have
\be v^M_0(x-y)=\sum_{k=0}^{j-1}   C^{M-k}( L^{-k}(x-y))  +
v^{M-j}_0(L^{-j}(x-y)) \ee
The  slice covariances are  defined by\footnote{ We  have chosen to take
$e^{-p^4}$
rather than say  $e^{-p^2}$ in (\ref{vdef}),(\ref{cm}) in order to have a
smoother approach  to infinite volume at $p=0$. }
\begin{equation}  \label{cm}
C^M(x-y)=|\La_M|^{-1}\sum_{p\in \La_M^*, p \neq 0}
\frac{e^{ip(x-y)}}{p^2}(e^{- p^4} -e^{-L^4p^4})
\end{equation}
The   integral over  $\mu_{\beta v^M_0}$ in the partition
function can then be evaluated by successively taking
convolutions with
$\mu_{\beta C}$ and then scaling down by  $L$.
After  $j$ steps we have the expression
\be
Z=\int \cZ_j(\f) \ d\mu_{\beta v^{M-j}_0}(\f) \label{Z1}
\ee
with successive densities $\cZ_j$ defined on $\La_{M-j}$
and  related by
\be \cZ_{j+1}(\f)= (\mu_{\beta C}*\cZ_j)(\f_L)
= \int \cZ_j(\phi_L + \z) d\mu_{\beta C}(\z)
   \label{RGtrans} \ee
where $\f_L(x)=\f(x/L)$
is the canonical rescaling of the field for $d=2$.
Equation (\ref{RGtrans}) is the RG map.   We want to study the flow starting
with  $\cZ_0(\phi)  =   \exp (\z \int \cos \phi )$.

To track the flow we must isolate the fastest growing parts of $\cZ_j$
during each  RG step.
 We extract a constant part
and a gradient part
 and  instead of (\ref{RGtrans}) now  define
$\cZ_{j+1}$ by   \be \cZ_{j+1}(\f)\  \exp \left(\de E_j |\La_{M-j}| -
\frac{ \de \sigma_j}
{2 \beta}  \int_{\La_{M-j-1}} (\pa \phi)^2 \right)
= \left(\mu_{\beta C}*\cZ_j\right)(\f_L)
   \label{RGtrans2} \ee
with special choices of   $\de E_j, \de \si_j$.
The quadratic factor is absorbed into the measure  at each step  and so
instead
of  (\ref{Z1}) we have  for some constants   $ \cE_j, \si_j$
\be
Z=  e^{\cE_j}\int \cZ_j(\f) \ d\mu_{\beta v^{M-j}_0(\sigma_j)}(\f) \label{Z2}
\ee
where
\begin{equation}
v^M_0(\si; x-y)=|\La_M|^{-1}\sum_{p\in \La_M^*, p \neq 0}
\frac{e^{ip(x-y)}}{p^2}(e^{p^4} +  \si)^{-1}
\end{equation}
The successive values of $\si_j$ are given by  $\si_{j+1}  = \si_j + \de \si_j$
and there is a similar formula for  $\cE_{j+1}$ in terms of  $\cE_j, \de E_j$
and $\de \si_j$.

To state the  main result we need one more ingredient. 
This is a local structure for  the densities $\cZ_j$.
Following Brydges and Yau \cite{BrYa90}
densities are represented  by polymer expansions
  $\cZ_j(\phi) = \cE xp ( \Box + K_j)(\La_j, \phi)$
 as we now explain.
 A {\it closed polymer} $X$ is a union of closed unit squares
centered  on  lattice
points.   A {\it polymer activity} is a function $K(X, \f)$ depending on
polymers $X$
and fields $\f$ with the property that the dependence on $\f$ is localized
in $X$.  One can define a product on polymer activities and an associated
exponential function  $(\cE xp (K))(X, \phi)$.   If  $\Box $ is the
characteristic
function of open unit cells then
\be \cE xp ( \Box + K)(X, \f) = \sum_{\{X_i\} } \prod_i K(X_i, \f) \ee
where the sum is over collections of disjoint polymers $\{ X_i\}$ in $X$.
For an exposition of polymers see \cite{BDH94a}.

Now we can state the IR result:

\bthm \label{irthm1} Let $\beta > 8 \pi$, let  $ \ep >0$, let  $L$ be
sufficiently large, and  let
$|\zeta|$ be  sufficiently small.   Then for  $j=0,1,2,..$ the partition
function $Z$ defined by (\ref{Z0}) with $N =0$ can be written
\be Z=e^{\cE_j}  \int \cE xp ( \Box    + K_j)(\La_{M-j},\f)
 d \mu_{\beta v^{M-j}_0(\si_j)}(\phi)
\ee
where $\cE_j/ |\La_M|$ and $ \si_j$ are bounded and $\cO(|\zeta|)$ uniformly in
$M$.
 The polymer activities $K_j$ are even and $2\pi$--periodic in $\f$.
There is a  norm $\| \cdot  \|_{\infty}$ such that
\be\|K_j\|_{\infty} \leq \de^j|\z|^{1-\ep} \ee
where  $\de= \one \max\{L^{-2},L^{2-\beta/4\pi}\}<1/4$
\ethm

Here and throughout the paper $\one$ means a constant which is 
independent of $L, \z, M,j$.
The norm $\|K_j \|_{\infty}$ of $K_j(X, \phi)$  enforces conditions
of growth and analyticity in $\phi$ and requires tree decay in
$X$. A  more precise version of the theorem will be stated later  when
we come to the proof.

The point is that $K_j$ shrinks uniformly in $M$ so that the dominant
contribution  as $ j \to M$ is from the Gaussian measure.
The result gives a uniform bound on the energy density  $\log Z/|\La_M|$
and it should be possible to also take the limit $M \to \infty$.
Everything should also be analytic in $\z$ in a complex neighborhood
of the origin.  The only difficult part here is working with
complex measures;  see  \cite{BrYa90} for a treatment of this
problem for the closely related dipole gas.

A modification of this theorem to include local perturbations
should make it possible to study correlation functions for
the model,   proving the existence of the $M \to \infty$
limit and showing that the long distance behavior of correlations
is essentially the same as free.
See  \cite{Hur95} for results of this nature,
and  \cite {DiHu92},  \cite{BrKe93}  for the closely related
dipole gas.

Let us mention some earlier work  on this model.   It was  first
treated heuristically by Kosterlitz and Thouless \cite{KoTh73}.
Fr\"{o}hlich and Spencer later gave a rigorous treatment for
$\beta $ large \cite{FrSp80} by a special method (not the RG).
The range of validity was extended to  $\beta > 8 \pi$ by
Marchetti and  Klein  \cite{MaKl91}.

\bigskip

Now we discuss the UV problem for
 $\beta  < 8 \pi $.
We start with a fixed torus $\La_M$.  For simplicity take
the unit torus $\La_0$ so the starting covariance is $v^0_{-N}$.
We also make a renormalization replacing
$\cos(\phi(x))$ by the Wick ordered version
\be   : \cos (\phi(x) ) :_{\beta v^0_{-N}}  =   \exp ( \beta v^0_{-N}(0)/2 )
\cos(\phi (x)) \ee
Thus we study the
 partition function
\be Z = \int  \exp  \left( \z \int_{\La_0} : \cos (\phi(x) ):_{\beta
v^0_{-N}} dx  \right)
 d \mu_{\beta v^0_{-N}}(\phi)  \label{Z4}  \ee
 We scale up  to get an expression for $Z$ on $\La_N$.  Absorbing
the Wick ordering constant into the coupling constant one finds
that
\be Z = \int  \exp  \left( \z_{-N} \int_{\La_N}  \cos (\phi(x) ) dx  \right)
 d \mu_{\beta v^N_0}(\phi) \label{Z5} \ee
where  for any $j\le 0$
\be   \z_{j } =
 L^{-2|j|}  \exp ( \beta v^0_{-|j|}(0)/2 ) \z  \approx
L^{-(2-\beta/4\pi)|j| }\z  \ee

 The UV problem of controlling the limit $N \to \infty$ by an RG
analysis looks
very much like the IR problem.   The main difference is that the coupling
constants
$\z_{j}$ start out ultra small at $j=-N$ and   grow to a small value
$\z_0=\z$,
instead of starting out   small and  then shrinking.
A technical simplification  is that the field
strength extraction is no longer needed and we can take $\si_j =0$.

We define
\be    V(X, \phi )   =
 \left\{
\begin{array}{rl}  \int_{\De} \cos (\phi(x)  ) dx  &  \ \ \ \ \ \
X=\De= {\rm
unit} \ {\rm square}  \\
0 & \ \ \  \ \ \  |X| \geq 2 \end{array}
\right.
\label{vdefn}\ee
and then the result is :

\bthm\label{uvthm}  Let $\beta < 8 \pi$, let $  \ep > 0 $, let $L$ be
sufficiently large, and  let
$|\zeta|$ be  sufficiently small.  Then for  $j=-N, -N+1,\dots,0$ the
partition function given by (\ref{Z4}) or (\ref{Z5})  can be
written
\be Z=e^{\cE_j}  \int \cE xp ( \Box  + K_j)(\La_{|j|},\f)
 d \mu_{\beta v^{|j|}_0}(\phi) \ee
where
\bea   \cE_j  & =&  \sum_{k = -N}^{j-1}  \de  E_k | \La_{|k|}| \nn \\
 K_j  &= & \z_j V +\tilde K_j  \eea
The  $ \tilde K_j$ are even and $2\pi$--periodic in $\f$, and
 $\de E_j,  \tilde K_j$ are analytic in $\z$ and satisfy
\bea   | \de E_j |  &  \leq  &   |\z_j|^{2-\ep} \nn \\
\| \tilde  K_j \|_{\infty}  & \leq &  |\z_j|^{2-\ep}  \eea
for some norm  $\|\cdot\|_{\infty}$ .
\ethm

For  $\beta < 4 \pi$
the theorem implies that $Z$ is  uniformly bounded
and  analytic in $\z$.  For
  $4\pi \leq \beta  < 8\pi$ it isolates the divergence in   $Z$.
 One can also show that   $\de E_j, K_j$  have limits as $N \to \infty$,
and hence so does  $Z$   (for   $\beta < 4 \pi$). \cite{DiHu93}

A modification of this theorem to include local perturbations
should make it possible to study correlation functions for
$\beta < 8 \pi$ .  (The potentially  divergent  factor  $\cE_j$ does not
contribute
to correlation functions). One should be able to take
the $N \to \infty$ limit  and study the short distance behavior
of correlations.
See \cite{DiHu93},\cite{Hur95} for results of this nature.
Also see \cite{Dim98} for a proof that at $\beta = 4\pi$
the theory is equivalent to a theory of massive free
fermions.

Earlier work on this problem can be found in \cite{FrSe76},
\cite{BGN82}, \cite {NRS86}
\cite{NP89}.

\section{Estimates on the  RG map}

Our treatment of the RG map on polymer activities is  similar
to  that used in previous papers \cite{DiHu91},\cite{DiHu93},
\cite{BDH95}, \cite{BDH98a}.
However there are essential modifications: 
references \cite{BDH95}, \cite{BDH98a},
which we follow as
much as possible, use open polymers while we have to use  closed polymers
as in
\cite{DiHu91},\cite{DiHu93} (see the discussion in the next
section ). Our norms are  now simpler
as well,  a simplification we pay for with some harder proofs.

In this chapter, we analyze a single RG map
on a torus  $\La = \La_M = \bR^d/ L^M \bZ^d$ of 
arbitrary dimension $d\ge 2$.
We work with the fixed covariance
\be   \label{Mcovariance}
 C^M(\si, x-y)=|\La_M|^{-1}\sum_{p\in \La_M^*, p \neq 0}
\frac{ e^{ipx}}
{p^2}\ [(e^{p^4}+\si)^{-1}-(e^{L^4p^4}+\si)^{-1}]
\ee
and $|\si|$ assumed small,  although the results holds
for a much larger class.
 We start by defining our norms. Then we
consider separately
the three pieces of the RG: fluctuation, extraction, and  scaling. Finally
we put them
together in Theorem \ref{rtheorem} to give an overall estimate on  the RG map.

\subsection{Norms}

Let the Banach spaces $\cC^r(X),\cC^s(X)$ of smooth fields $\phi(x)$ on a
closed polymer $X$
be defined respectively for fixed $r,s\ge 0$ by the following norms: \bea
\| \f\|_{X}\equiv \| \f\|_{\infty, r,X} &=&    \sup_{|\al | \leq r,
\ x \in
X}   |\pa ^{\al}\f(x)|  \nn\\
\| \f\|_{s,X}\equiv  \| \f\|_{2,s,X} &=&   \left[ \sum_{|\al | \leq s}
\int_X |\pa
^{\al}\f(x)|^2\ dx\right]^{1/2}
\label{firstnorm}\eea
We assume $s>d/2 +r$ to ensure a Sobolev inequality $\| \f\|_{\infty,
r,X}\le \one  \|
\f\|_{2,s,X} $ and the corresponding dense embedding $\cC^s(X)\subset
\cC^r(X)$.

Let $K(X,\phi)$ be a smooth  function on $\cC^s(X)$.   Thus we assume the
existence  of all derivatives  $K_n(X,\f)$.  These are  
continuous  symmetric multilinear
functionals on $\cC^s(X)$. In fact  make  
 stronger assumption that these derivatives have continuous extensions
to  $\cC^r$ by demanding the finiteness of the following norm
\be
\label{varnorm}
\|K_n(X, \f) \| =  \sup_{f_i \in \cC^s(X)\atop\|f_i\|_{\infty,r,X} \leq 1}
|K_n(X,
\f;f_1,...,f_n) | .\ee

A {\it large field
regulator} is a functional of the form
\be G( \k,X, \phi ) =
 G'(\k, X, \phi ) \de G(\k, \pa X, \phi )   \label{gee}
\ee
where
\bea   G'(\k, X, \phi )
&=&   \exp  ( \kappa    \sum_{1 \leq | \al| \leq s}
\int_X  |\pa^{\al} \phi |^2   )  \nn \\
 \de G(\k, \pa X, \phi )
&=&   \exp  ( \kappa c  \sum_{ | \al |=1}
\int_{\pa X}  |\pa^{\al} \phi |^2  )
\eea
with constants $\k, c \leq 1$  to be specified.

 A {\it large set regulator} $\G(X)$
has the form
\be
\G(X) = A^{|X|} \Th(X)  \label{gamma}
\ee
for a parameter  $A\ge 1$ and  factor $\Th(X)$  such that
$\Th(X)^{-1}$ has polynomial tree decay  (see
\cite{BrYa90},
\cite {BDH98a} for the exact definition). For our present paper we fix
$A=L^{d+3}$, and
also define
$\G_p(X)=2^{p|X|}\G(X)$ for any
$p=\pm1,\pm 2,\dots$.

 In terms of regulators $G,\G$ and a further parameter $h\ge 0$ we
define the norms:
\bea
\|K\|_{G,h,\G}  &=&   \sum_{X\supset
\De}\G(X)\|K(X)\|_{G,h}  \nn \\
\|K(X)\|_{G,h}&=&\sum_{n=0}^{\infty}\frac{h^n}{n!}
\|K_{n}(X)\|_{G}  \nn\\
\|K_{n}(X)\|_{G}&=&
 \sup_{\f \in \cC^s(X)}\|K_{n}(X,\f) \|G(X,\f)^{-1}.
\label{gbound}
\eea
The sum over $X$ is independent of the unit block $\De$ for translation
invariant $K$.

These norms are simpler than the norms in earlier versions of this
formalism in which one first localizes the derivatives in unit blocks, then
takes the supremum
over the fields, and finally sums over blocks.   The previous version
(designed for
models in $d>2$) controls the fluctuation step in an elegant manner, but in
$d=2$ leads to
unbounded  growth in the parameter $h$ in the scaling step. The present norms
require a different treatment of
fluctuation, but avoid growth in $h$.

Another point concerns  the boundary term  $\de  G( \k,\pa X, \phi)$ in the
large
field regulator  $  G( \k, X, \phi)$.
It is present  to absorb the growth of  $G'(\k, X, \phi)$
a feature upon which  we elaborate in the next section.
However we also need
$G(X)G(Y) \leq   G(X \cup Y)$ for disjoint
polymers.  The boundary term spoils this if the polymers
are open since disjoint polymers may have pieces of their
boundaries in common.  This is the reason we have taken closed
polymers.

\subsection{Fluctuation}
Given a localized density   $\cE xp ( \Box + K)$ and the Gaussian measure
$\mu_C$
we want to find new polymer activities $\cF K$  such that
$  \mu_{C}  * \cE xp (  \Box + K  )  =  \cE xp ( \Box + \cF K)$
and such that we more or less preserve control over  size and
localization.  We accomplish this using the framework of Brydges and Yau
\cite{BrYa90}
(see also Brydges and Kennedy  \cite{BrKe87}).
Those authors actually give two constructions.   The first is by solving
a  functional Hamilton-Jacobi equation for  $  \mu_{tC}  * \cE xp (  \Box +
K  )$.
This is elegant and efficient, but with our new norms we cannot take
advantage of it
(since we can no longer keep the large set  regulator $\G$ constant). Instead
we use the second construction of Brydges and Yau,
an explicit  cluster expansion.  Cluster expansions have a long history
starting with  \cite{GJS73}.

We begin with the purely combinatoric part.
Let $F(s)$ be a continuously differentiable function of $s= \{ s_{ij} \} $
where
$0 \leq s_{ij} \leq 1$ and
where  $ij$ runs over the distinct unordered pairs  ({\it bonds})  from
some finite index
set.    A {\it graph}  $G$ on this set is a collection of bonds , and it  is
called  a {\it forest} if it has no closed loops.  The set of all forests is
denoted  $\cF$.  Finally we define
\begin{equation}  \label{ar}
\sigma_{ij} (G,s)  =   \inf \{  s_b : b \in {\rm path}  \ {\rm joining}\ ij \
{\rm in}\ G \}
\end{equation}
with the convention that $\sigma_{ij} (G,s) =0 $ if there is no path joining
$ij$ in $G$.
Then for $1=\{1,1,...,1\} $ 
\begin{equation}  \label{tree}
F( 1 )  =  \sum_{G\in \cF} \int\ \prod_{b\in G} ds_b\ (\prod_{b\in G}
\pa_{s_b} F) (\sigma(G,s))
\end{equation}
where the  $G= \emptyset$ term is interpreted
as   $F(0)$.
For the proof see  Abdesalam and Rivasseau  \cite{AbRi95}
or Brydges and Martin, Theorem VIII.2  \cite{BrMa99}.

 Now for any $X,Y$ define
\begin{equation}
C(X,Y)(x,y) = \frac12[ \chi_{X}(x) \chi_{Y}(y)+\chi_{Y}(x) \chi_{X}(y)]  C(x-y)
\end{equation}
and let   $C_X =  C(X,X)$ be the restriction of $C$ to $X$.
Suppose that $\{X_i\}$  is  a collection of disjoint polymers whose
union
is $X$.
Then the restriction  $C_X$ can be written
\be C_X =   \sum_{i,j}  C(X_i,X_j) \ee
with the sum over ordered pairs.
We weaken the coupling between  $X_i, X_j$ with  parameters $s_{ij}$
and define
\begin{equation}  \label{csx}
C_X(s) =   \sum_{i,j}  C(X_i,X_j)  s_{ij}
\end{equation}
where $s_{ii} =1$.
Now while $C_X(s)$ is not necessarily positive definite,
  $C_X( \sigma(T,s)) $
 is  positive definite
for any $s$ and any  tree $T$ \cite{BrKe87}, \cite{BrMa99}.

Let
$ \De_{C}$ be the functional Laplacian given formally by:
\begin{equation}
\De_C   = \frac12 \int C(x,y)  \frac {\de}{\de  \phi (x)}
\frac {\de}{\de  \phi (y)}  dx dy
\end{equation}

\begin{lemma}  \cite{BrYa90},  \cite{AbRi95}
     For smooth polymer activities  $K$
\be \mu_{C}  * \cE xp (  \Box + K  )  =  \cE xp ( \Box + \cF K)  \ee
with
\be   \label{kaytee}
 \cF K(X)  = \sum_{ \{ X_i \},T \to X }
  \int ds^T \  \mu_{C_X(\sigma(T,s))} * \prod_{ij \in T}
(-2\De_{C(X_i,X_j)} ) \prod_i K(X_i)
\ee
where  the sum is over collections of  disjoint  polymers  $\{ X_i \} $
 whose union is $X$, and over
tree graphs  $T$ on  $\{ X_i \} $.  If $\{ X_i \} = \{ X\} $ the term is
interpreted
as  $ \mu_{C}* K(X)$.
\end{lemma}

\pr  We start with
\begin{equation}
    \mu_{C}  * \cE xp (  \Box + K  ) (X)
=   \sum_{ \{ X_i \} } \mu_{C}  *
 \prod_{i} K(X_i)
\end{equation}
In the expression   $ \mu_{C}*\prod_{i} K(X_i,\phi)$
we regard the product as a function of fields
$\phi$ on $X$ only, and replace the covariance $C$ by $C_X$.\footnote{Our
assumption that   $K(X, \phi)$  has $\phi$  dependence localized in  $X$
means that  the function is  measurable with respect to  $\Sigma_X$, the
$\sigma$-algebra  generated by  $\{ \phi_x \}_{x \in X}$ .  }
If $\{ X_i \}=\{ X \}$ has only one element we leave the
expression alone.  Otherwise there are two or more
subsets and  for each $\{ X_i \}$ we analyze  $F(1) = \mu_{C_X}  *
 \prod_{i} K(X_i)$
by introducing  the interpolation
 $F(s) = \mu_{C_X(s)} *
 \prod_{i} K(X_i)$  with  $C_X(s)$ given by (\ref{csx}), and then
making the expansion (\ref{tree}).
This  gives the expression
\begin{equation}
  \sum_{ \{ X_i \} }
 \sum_G  \int ds^G \left( \pa^G \mu_{C_X(s)} *   \prod_{i}K(X_i)
\right)_{s=\sigma(G,s)}
\end{equation}
Now the graph $G$ can be regarded as a union of trees  $\{ T_k\}$.
Grouping together the polymers  $\{ X_i \} $ linked by  the trees
yields  new disjoint polymers $\{ Y_{k} \} $.  The covariance $C_X(s)$
preserves
the  $\{ Y_{k} \} $ since $\si_{ij}(G,s) = 0$ for blocks $X_i,X_j$   in
different trees.    We can write   $C_X(s)  =  \oplus_{k} C_{Y_k}(s) $.
Then the
integrand above factors and we have
\begin{equation}
  \sum_{ \{ X_i \} }
 \sum_{ \{T_k \} }   \prod_k \left[ \int ds^{T_k}
 \left( \pa^{T_k} \mu_{C_{Y_k}(s)} *   \prod_{i:X_i \subset Y_k} K(X_i)
\right)_{s=\sigma(T_k,s)} \right]
\end{equation}
 Now we group together the terms in the sum by the $\{ Y_k \}$ they
determine and
find
\begin{equation}
  \sum_{ \{Y_k \} }
 \prod_{k} \cF K(Y_k)  = \cE xp (  \Box + \cF K  )
\end{equation}
where
\begin{equation}  \cF K(Y)  =  \sum_{ \{X_i\}, T  \to  Y}
  \int  ds^{T} \left( \pa^{T} \mu_{C_Y(s)} *   \prod_{i} K(X_i)
\right)_{s=\sigma(T,s)}\end{equation}
The result now follows since
 $\pa C_Y(s)/ \pa s_{ij}  =  2 C(X_i, X_j) $ and hence
$\pa / \pa s_{ij}( \mu_{C_Y(s)} * F) =    \mu_{C_Y(s)} *(- 2\De_{C(X_i,
X_j)}) F$
 and hence
 \begin{equation}
\pa^{T} \mu_{C_{Y}(s)} *   \prod_{i} K(X_i)
= \mu_{C_{Y}(s)} *  \prod_{ij \in T}(-2 \De_{C(X_i,X_j)} )  \prod_{i} K(X_i)
\end{equation}

\QED
\bigskip

The behavior in $\phi$ of $\cF K(X, \phi)$ will
turn out to be slightly worse than that for  $K(X,\phi)$ which means we
have to take a larger  large
field regulator. It is convenient to choose a regulator which is a scaling
of the
original.   Let the field scaled up by  $ \ell >1$ be defined by
\be   \phi_{\ell}(x) =  \ell^{-(d-2)/2}\phi (x/\ell)  \ee
(our convention here is different from  earlier papers).
Then define
\bea  G_\ell(\k ,X, \phi ) &=&  G(\k, \ell^{-1}X, \phi_{\ell^{-1}}) \nn \\
&=&   \exp  \left( \kappa    \sum_{1 \leq | \al| \leq s}  \ell^{2 |\al| -2}
\int_X  |\pa^{\al} \phi |^2  +   \kappa c  \sum_{ | \al |=1}
 \ell  \int_{\pa X}  |\pa^{\al} \phi |^2  \right)  \label{scaledG}
\eea
For the applications we have in mind we need  $1 < \ell \leq L$: for
definiteness we take  $\ell=2$ in the following.

For unit blocks  $\De, \De'$ define
\be  C_{* }(\De, \De')  = \|C(\De, \De')\| d(\De, \De')^{2d}
 \theta (\De, \De')
\ee
Here the norm is the  $\cC^r$ norm in each variable,  $d(\De, \De')$ is
Euclidean distance between block centres, and
$\th$ is the distance
function built into the tree decay factor $\Th$ in (\ref{gamma}). Now define
\be   \|C\|_{*} =\sup_{\De}  \sum_{\De'\ne\De}
C_{*}(\De, \De')   \ee

\bthm  \label{flucthm}  Let   $\k c^{-1} L^2$ be sufficiently small.
 Then there is a constant $\gamma$ depending
only on the dimension such that if  $0 <\de h < h$ and for some $p$
\be\label{deltah}  \de h^2  \geq 8 \g^2\ \| C\|_{*}\ \| K\|_{G(\k), h ,
\G_{p+3}}
\ee
then 
\be  \|  \cF K \|_{G_{\ell}(\k),h - \de h ,\G_{p} }
 \leq  2\| K\|_{G(\k),h, \G_{p+3} }
\ee
\ethm
\bigskip

\re
The linearization   $\cF_1 K  = \mu_C * K$ satisfies the same bound
(or  even the better bound with $\de h =0$).
\bigskip

\pr   We adapt the analysis of \cite{BrYa90} to our norms.
In  (\ref{kaytee}) change to a sum on disjoint ordered polymers
$(X_1, ..., X_N)$  and regard  $T$ as a tree on  $(1,...,N)$.
Then we have
 \bea
\cF K(X) &=& \mu_{C}* K(X)   \nn \\
&+&  \sum_{N=2}^{\infty}\frac{1}{N!} \sum_{(X_1, ...,X_N)}
 \sum_T  \int ds^T  \mu_{C_X(\sigma(T,s))} * \prod_{ij \in T}
\De_{C (X_{i},X_{j})}  \prod_{i=1}^N K(X_i)
\eea
We next introduce a sum over unit blocks: for  $b=\{ij\}$
\be   \De_{C (X_{i},X_{j})}
=   \sum_{\De_{bi} \in X_{i},\De_{bj} \in X_{j}}
  \De_{C (\De_{bi},\De_{bj})}
\ee
Taking derivatives and  norms yields
\bea
&& \|(\cF K)_n(X)\|  \leq  \| \mu_{C}* K_n(X) \|
\ + \ \sum_{N=2}^{\infty}\frac{1}{N!} \sum_{(X_1, ...,X_N)}
 \sum_T  \sum_{ \{ \De_{bi}, \De_{bj}  \} }  \nn \\ &&
\sum_{n_1,...,n_N} \frac {n!}{n_1!...n_N!}
\int ds^T \|  \mu_{C_X(\sigma(T,s))} *\left[ \prod_{b \in T}
\De_{C (\De_{bi},\De_{bj})}  \prod_{i=1}^N K_{n_i}(X_i)  \right] \|
\eea

In  Lemma   \ref{CX} to follow we show that  for $\k c^{-1} L^2$
sufficiently small
\be    \mu_{C_X(\sigma(T,s))} *  G(\k, X)  \leq  G_{\ell}(\k ,X)2^{|X|} \ee
which makes it possible to estimate the above convolutions.  Using this and
  $G(\k,X) =  \prod_i G(\k,X_i)$ (since the  $X_i$ are disjoint)
we find  (see \cite{BDH98a} for more details)
\bea
&& \|(\cF K)_n(X)\|_{G_{\ell}(\k)} \leq \|  K_n \|_{G(\k)}
\ + \ \sum_{N=2}^{\infty}\frac{1}{N!}
 \sum_{(X_1, ...,X_N)}
 \sum_T   \sum_{ \{ \De_{bi}, \De_{bj}  \} } \nn \\ &&
 \sum_{n_1,...,n_N} \frac {n!}{n_1!...n_N!}
  \prod_{b \in T}
\|C (\De_{bi},\De_{bj}) \| \prod_{i=1}^N  \| K_{n_i+ d_i}(X_i)
\|_{G(\k)} \ 2^{|X_i|}
\eea
Here  $d_i$ is the incidence number for the $i^{th}$ vertex in the
graph   $T$.

Now multiply by $(h- \de h)^n/n!$ and sum over $n$ to obtain
\bea
 \|\cF K(X)\|_{G_{\ell}(\k),h-\de h} &\leq& \|  K_n(X) \|_{G(\k), h}
\ + \ \sum_{N=2}^{\infty}\frac{1}{N!}
 \sum_{(X_1, ...,X_N)}
 \sum_T   \sum_{ \{ \De_{bi}, \De_{bj}  \} } \nn \\ &&
\hspace{-.3in}\prod_{b \in T}
\|C (\De_{bi},\De_{bj})\| \prod_{i=1}^N\Bigl(\frac{d}{d
h}\Bigr)^{d_i}
 \| K(X_i)\|_{G(\k), h - \de h} \ 2^{|X_i|}
\eea
A  Cauchy  bound yields
\be(\frac{d}{dh})^{d_i}   \| K(X_i) \|_{G(\k), h-\de h}
\leq \Bigl(\de h\Bigr)^{-d_i}  d_i! \| K(X_i) \|_{G(\k), h}  \ee
It is proved in Lemma \ref{factbd} to follow that for any $i$ we have
\be    d_i! \leq  \g^{d_i}  \prod_{b \ni i}
 d(\De_{bi}, \De_{bj})^d \label{bound} \ee
for some constant $\g$.   Taking into account that   $\sum_i d_i = 2N-2 $
this gives
\be   \prod_i d_i!  \leq  \g^{2N-2} \prod_b d(\De_{bi}, \De_{bj})^{2d} \ee
and so
\bea
&& \|\cF K(X)\|_{G_{\ell}(\k),h-\de h} \leq  \|  K(X) \|_{G(\k), h} +
 \sum_{N=2}^{\infty}\frac{1}{N!}
 \sum_{(X_1, ...,X_N)}
 \sum_T   \sum_{ \{ \De_{bi}, \De_{bj}  \} } \nn \\ &&
(\g\ \de h^{-1})^{2N-2}\prod_{b \in T} 
\|C(\De_{bi},\De_{bj})\|_{\infty} d(\De_{bi}, \De_{bj})^{2d}
\prod_{i=1}^N
 \| K(X_i)\|_{G(\k), h}\ 2^{|X_i|}
\eea

Now multiply by
\be\label{gamma1}  \G_{p}(X) \leq  \prod_i \G_{p}(X_i)
\prod_{b}  \theta (\De_{bi},\De_{bj}) \ee
and identify   $ \prod_b C_{*}(\De_{bi}, \De_{bj})$.
Next  sum over  $X \supset \De$ and dominate the expression by 
a sum over $i_0$  and a sum over unrestricted disjoint
$(X_1,...,X_N)$ such that $X_{i_0} \supset \De$.
To estimate this sum  and the sum over
$\{ \De_{bi}, \De_{bj}  \}$, we start at the twigs of the tree and work
inward leaving to the  last
the set $X_{i_0}$ which is pinned.  Suppose that when
we come to a
vertex
$i$ we have gained a factor
$|X_i|^{d_i -1}$  from the previous estimates.  If $b=\{ij\}$
is the remaining inward bond at this vertex and   $\De = \De_{bi}, \De' =
\De_{bj} $, then we
have
\bea  &&  \sum_{X_i} \sum_{\De \in X_i,\De' \in X_j}
C_*(\De,\De')
 \| K(X_i) \|_{G(\k),h} \G_{p+1}(X_i) |X_i|^{d_i-1}  \nn \\
& \leq &  \sum_{X_i} \sum_{\De \in X_i,\De' \in X_j}
C_* (\De,\De')
 \| K(X_i) \|_{G(\k),h} \G_{p+3}(X_i) (d_i-1)!  \nn \\
& \leq &
 \sum_{\De' \in X_j,\De}
C_* (\De,\De')
 \| K \|_{G(\k),h, \G_{p+3}}  (d_i-1)!  \nn \\
& \leq &
 \|C \|_{*}
 \| K \|_{G(\k),h, \G_{p+3}} |X_j| (d_i-1)!  \eea
This gives a factor  $|X_j| $ for the $j$ vertex.
 The  case for $i=i_0$ is special and we have
$d_{i_0}!\le (N-1)
(d_{i_0}-1)!$) .  There is also a factor  $N$ for the sum over  $i_0$
and combining all the above yields
\be
\|\cF K\|_{G_{\ell}(\k),h- \de h,\G_{p}} \leq \|  K \|_{G(\k),h, \G_{p+3}}
 (1+  \sum_{N=2}^{\infty}\frac{\al^{N-1}}{(N-2)!}
 \sum_T
  \prod_{i=1}^N (d_i-1)!  )
 \ee
where  $\al =
\g^2 \ \de h^{-2}\ \| C \|_{*}  \| K \|_{G(\k), h,\G_{p+3}}  $.
But
the number of trees with given incidence numbers  $d_i$ is
$(N-2)!/ \prod_i (d_i-1)! $ by Cayley's theorem, and the number
of choices of $d_i$ is bounded by $2^{2N-2}=4^{N-1}$.  Thus the sum over
$T$ is bounded by   $(N-2)! 4^{N-1}$.   
Then the sum over  $N$ is bounded by   $\sum_{N=2}^{\infty} (4 \al)^{N-1}$
and this is  less that $1$  since our basic assumption is  $4 \al \leq 1/2$

\QED
\bigskip

This completes the proof of the theorem, except for the following two
results  which we skipped.

\blem
\label{CX}
 Let   $\k c^{-1} L^2$  be sufficiently small.
Then
\be   \mu_{C_X(\si(T,s))} * G(\k,X)  \leq  G_{\ell}(\k ,X)2^{|X|}
\ee
\elem

\pr  (see  \cite{BDH94a} for more details)   Consider for  $0 \leq t \leq
1$ the family of large field regulators
\be\label{Gt} G_{t}(\k, X)= 2^{t|X|}\left[ G_{\ell}(\k,X)\right]^t\left
[G(\k,X)\right]^{1-t}
\ee
We prove for  $0 \leq t \leq 1$  that
\be   \mu_{tC_X(\si(T,s))} * G_0(\k,X)  \leq  G_t(\k ,X)  \label{homotopy}
\ee
The result we want comes at  $t=1$.

 We have  $G_t(X) = \exp(U(t,X))$  where (with $\ell=2$)
\begin{equation}
  U(t,X)  = t \log(2) |X|
     + \kappa \sum_{1 \leq |\alpha| \leq  s}
       \int_{X} |\pa^\alpha \phi|^2
       \cdot  ( 2^{2|\al| -2}t +(1-t)  )
+ \k c \int_{\pa X} | \pa \phi |^2  (1+t)
  \label{U}
\end{equation}
The bound (\ref{homotopy})
is implied by
\begin{equation}
\Delta_{C_X(\si(T,s))} U   +  {1 \over 2}
C_X(\si(T,s)) \left({\pa U\over \pa \phi},{\pa U\over \pa \phi}\right)
 \leq   {\pa U\over  \pa t}
  \label{R9}
\end{equation}
Showing (\ref{R9}) is a somewhat lengthy computation in which every term on
the left is bounded by corresponding terms on the right for
$\k$ sufficiently small.
The terms with $|\al| =1$ are special since there is no corresponding term
on the right.   Instead one integrates by parts. This  adds derivatives
and boundary terms both of which can be bounded.

The condition on  $\k$ turns out to be  that the following
quantities be sufficiently small:
\bea  && \k \
  \sup_{1 \leq |\al|, |\beta| \leq s} \sup_{x \in X}| ( \pa_x ^{\al}
\pa_y^{\beta} C_X(\si(T,s)))(x,x) | \nn \\
 &&
\k c^{-1} \  \sup_{0 \leq |\al|, |\beta| \leq s} \sup_{x \in X}
 \int_{X} |( \pa_x^{\al} \pa_y^{\beta} C_X(\si(T,s)))(x,y) |dy  \nn \\
&&
\k c^{-1} \  \sup_{0 \leq |\al|, |\beta| \leq s} \sup_{x \in X}
 \int_{ \pa X} |( \pa_x^{\al} \pa_y^{\beta} C_X(\si(T,s)))(x,y) |dy
\label{kappabound}\eea
These quantities are bounded by the corresponding quantities with $\si=1$.
Note from Lemma
 \ref{cbound} in the appendix,
$(\pa_x^{\al}
\pa_y^{\beta}C)(x,x)$ is bounded by
$\one$.  The second and third quantities are bounded by same 
expressions with
$X=  \La $ and $X=$ the  $d-1$ dimensional ``checkerboard'' in $\La$.
 For both these
integrals, we use
 Lemma \ref{cbound} again and find the worst bound is   $\k c^{-1} L^2$ .   Hence
the  result follows.

\QED

\bigskip

\blem\label{factbd}   Let $\De$ and  $\De_1, ... \De_n $ be distinct unit
blocks.  Then there is a constant $\g$ depending only on the
dimension $d$ such that
\be  n ! \leq   \g^{n} \prod_{j=1}^n d(\De, \De_j )^{d} \ee
\elem

\re  Bounds of this type were introduced
in  \cite{GJS73}.

\bigskip

\pr
  Let $m_r$ be the number of unit blocks intersecting a ball
of radius $r$ centered on a lattice point, and
select  $\g$ so $m_r \leq \g r^d$ for all $r>1$.
Order the blocks  so that
\be   d(\De, \De_{1} )  \leq  ... \leq  d(\De, \De_{{n}}) \ee
 Then  the ball of radius $r_k = d(\De, \De_{k})$
around the center of  $\De$ intersects $m_{r_k}$ unit blocks and  $m_{r_k}
\geq k$.
Then   $k \leq m_{r_k}  \leq  \g r_k^d$
and we have
\be  n!  = \prod_{k=1}^n k \leq   \prod_{k=1}^n  \g r_k^d
= \g^n  \prod_{k=1}^n  d(\De, \De_{k})^d
\ee

\QED
\bigskip

\subsection{Extraction}
In the extraction step we remove a polymer activity  $F$
from the general activity  $K$.  Usually $F$ is some low order terms in
$K$ but we do not assume this at first.
The extraction
 is defined so that
\be \cE xp(\Box + K)(\La,\f)= \exp \left(     \sum_{X \subset  \La} F(X,
\phi)\right) \
 \cE xp(\Box + \cE(K,F))(\La,\f)  \label{ex1}
\ee
with new polymer activities $ \cE(K,F)$.
To specify $\cE (K,F) $  we define
\bea
\tilde{K}(X) & =& K(X) - (e^{F} - 1)^+(X)  \nn \\
 (e^{F} - 1)^+(Y)  & =&   \sum _{ \{Y_j \} \to Y}  \prod_j (e^{F(Y_j)} -1)
\eea
where  the sum is over
collections $ \{Y_j \}$ of distinct polymers which are overlap connected
and  whose union is $Y$.
 Then formula
(\ref{ex1}) holds with  $\cE (K,F)$  given by
\be \label{Edef}
        {\cE}(K,F)(Z) =
        \sum_{\{X_i\},\{Y_j\}
        \to Z} \\
        \prod_i  \tilde K(X_i)
        \prod_j(e^{- F(Y_j)}-1).
\ee
where the  sum is over collections of disjoint
subsets $\{X_i\}$ and collections of distinct subsets $\{Y_j\}$
 each intersecting some $X_i$, so
that the   $\{X_i\}, \{Y_j\}$  are overlap connected
and their union  is $Z$.
 This version of extraction is taken from  \cite{DiHu92}, to 
which we refer for a proof.
The  linearization of  $\cE (K,F)$ in $K$ and $F$
 is    $\cE_1 (K,F)  = K -F $: this is the sense in which $F$ has been
removed from $K$.
\bigskip

To obtain estimates on $\cE(K,F)$ we will need estimates
like   $G(X) \leq  G(Z)$  when  $X \subset Z$.  For this to be
true we have to be able to dominate  $\de G$ by $G'$ so we can
``dissolve'' the pieces of $\pa X$ which do not contribute to $\pa Z$.
This means that the constant  $c$ in  $\de G$ has to be sufficiently small.
Let   $c_s$ be the Sobolev  constant  defined so that for $x \in  \De$,
the closed unit block, we have
$|\pa \phi(x)|^2 \leq
 c_s  \sum _{ 1 \leq |\al| \leq s} \int_{\De} | \pa^{\al}\phi |^2$.

\blem \label{disolve} For $X \subset Z$, $\kappa > 0$  and  $c <
(2d\ c_s)^{-1}$
we have
\be
G(\k, X)   \leq  G(\k,Z)
\ee
If   $c < (4d\ c_s)^{-1}$ the same bound holds
 with  $G$ replaced by $G_{\ell}$,
$ \ell=2$.
\elem

\pr  Let $f$ be a  face ($d-1$ cell) in
$\pa X$  which does not contribute to $\pa Z$.
Any such face $f$ must be also be a face for some $\De $ in $Z-X$.
Then we can ``dissolve'' the boundary by using the Sobolev inequality and
the bound on $c$ to obtain
\be  \de G (\kappa, f) \leq   G'( \kappa/2d, \De)
\ee
 Each $\De$  arises from at most $2d$ faces and
so
\be  \de G(\kappa, \pa X - \pa Z)
\leq    G'( \kappa, Z-X )
\ee
Thus we have
\bea   G(\kappa,X)  & = &  G'(\kappa, X) \de G( \kappa, \pa X - \pa Z )
 \de G( \kappa, \pa Z \cap \pa X ) \nn \\
&\leq &  G'(\k,Z) G( \kappa, \pa Z \cap \pa X )
 \eea
Since
$ \de G( \kappa, \pa Z \cap \pa X ) \leq    \de G( \kappa, \pa Z  ) $
the result follows.

\QED
\bigskip

We now    assume $F$ satisfies the following {\em localization\/}  property:
$F(X,\f)$ has the
decomposition
\begin{equation} \label{loc}
        F(X,\f) = \sum _{\Delta \subset X} F(X,\Delta ,\phi)
\end{equation}
where  $\De$ is summed over unit blocks, and $ F(X,\Delta ,\phi)$ has
the $\f$ dependence localized in $\De$.

We also  need   stability  conditions on
the perturbation $F$.  Let  $f(X)$ be a collection of constants.
 We say that $F$ is {\it stable} for
$(G, h, f(X))$ if for complex  $z(X)$
\begin{equation} \label{ex.2}
       \sup_{|z(X)|f(X) \leq 1} \|\exp  \left\{  \sum _{X \supset \Delta}
z(X) F(X,\Delta)
        \right\} \|_{G,h} \le 2
\end{equation}
For a method to verify the stability
hypothesis see the appendix.

\bthm  \label{exthm1}  Let  $c < (2d\ c_s)^{-1}$.
Suppose that  $F$ is  stable for
$(G(\k),h, f(X))$  and for  $(G'(\de \k),h,\de f(X))$    and that
$\|f \|_{ \G_{p+4}}$,$\|\de f \|_{ \G_{p+2}}$ and $ \|K\|_{G(\k),h, \G_{p+2}}$
are sufficiently small.
Then
there is a constant  $\cO (1)$
such that
\be
\label{ex4.1}
        \|\cE( K,F)\|_{G(\k+\de\k),h,\G_p} \leq
        \cO (1) ( \|K\|_{G(\k),h,\Gamma_{p+2}} +
        \|f\|_{ \Gamma_{p+4}} )
\ee
For   $c < (4d\ c_s)^{-1}$ the same bound holds with each  $G$ replaced by
$G_{\ell}$,
$\ell=2$.
\ethm
\bigskip

\pr The proof is similar to \cite{BDH98a} where however the extraction
is not global.  We start with
(\ref{Edef}) which can be written
\be
      {\cE}(K,F)(Z) =
        \sum_{\{X_i\},\{Y_j\} \to Z}
        \prod_i \tilde K(X_i)
        \prod_j \frac{1}{2 \pi i}\int \frac{dz_{j}}{z_{j}(z_{j}-1)}
        \exp
        \left\{
        - z_{j} F(Y_{j})
        \right\}
\ee
 The integral is over the circles $|z_j|\de f(Y_j)= 1$.
Inserting   $F(Y) = \sum _{\De \subset Y} F(Y, \De)$
we can rewrite this as
\begin{eqnarray}
     &&   {\cE}(K,F)(Z) \nn \\
&=&
        \sum_{\{X_i\},\{Y_j\} \to Z} \prod_i \tilde K(X_i)
        \prod_j \frac{1}{2 \pi i}\int \frac{dz_{j}}{z_{j}(z_{j}-1)}
        \prod_ {\De  \subset Z}
      \exp   \left\{
        -   \sum_jz_{j} F(Y_{j}, \De)
        \right\}
\end{eqnarray}

Now we note
\be   \prod_i G(\k, X_i)  \prod_{\De \subset Z}  G'(\de \k, \De)
 \leq G( \k + \de \k, Z)
 \ee
This follows from  $ \prod_i G(\k, X_i) = G(\k ,\cup_iX_i)  \leq  G(\k,Z) $
(by the lemma)
and from   $ \prod_{\De \subset Z} G'(\de \k, \De) = G'(\de \k ,Z)  \leq
G(\de \k,Z) $.
Using this estimate
and the multiplicative property of the norm we  obtain
\begin{eqnarray}\label{eintermed}
        \| {\cE}(K,F)(Z)\|_{G(\k+\de\k),h}
&\le&
        \sum_{\{X_i\},\{Y_j\} \to Z}
        \prod_i \|  \tilde K(X_i)\|_{G(\k),h}
        \prod_j \cO(1) \de f (Y_j)
\nn \\ &&
    \sup_{|z_j| \de f(Y_j) \leq 1} \prod_{\De \subset Z}
        \| \exp
        \left\{ - \sum _{j} z_{j} F(Y_j, \De)
        \right\}
        \|_{G'(\de \k),h}
\end{eqnarray}
By our second stability assumption the last factor is bounded by   $2^{|Z|}$.
Now  we write
\[\sum_{\{X_i\},\{Y_j\}}=\sum_{N,M}\frac{1}{N!M!}
\sum_{(X_1,\dots,X_N), (Y_1,\dots,Y_M)}\]
where the sum is over ordered sets, but otherwise the restrictions
apply.
We multiply by $\G_p(Z)$, identify $2^{|Z|}\G_p(Z) = \G_{p+1}(Z)$  and use
$ \G_{p+1}(Z) \leq \prod_i\G_{p+1}(X_i)
\prod_j\G_{p+1}(Y_j) $
which follows  from the overlap connectedness.
Then sum over $Z$ with a   pin, and use a spanning tree
argument and the small norm hypotheses to obtain
\begin{eqnarray} \label{ex10}
        \|{\cE}(K,F)\|_{G(\k+\de\k),h,\Gamma_p}
&\le&
        \sum_{N \ge 1, M \geq 0} \frac{(N+M)!}{N!M!}
        (\cO(1))^{N+M}
        \|\tilde K\|^N_{G(\k),h,\Gamma_{p+2}}
        \|\de f\|_{\Gamma_{p+2}} ^M \nn \\
&\le&
        \cO (1) \|\tilde K\|_{G(\k+\de\k),h,\Gamma_{p+2} }
 \end{eqnarray}
(In the last step use $(N+M)!/N!M!  \leq 2^{N+M}$ . )

Recall that $\tilde K=K- (e^{F}-1)^+$.   We write
\be
        (e^{F}-1)^+(Y)
= \sum_{\{Y_j\}}\prod_j \frac{1}{2 \pi i}
        \int \frac{dz_{j}}{z_{j}(z_{j}-1)}
        \exp
        \left\{  z_{j} F(Y_{j})
        \right\}
\ee
 now with the integral over  $|z_j| f(Y_j) = 1$.
Proceeding as above and using the first stability
assumption we have
\be  \| (e^{F} -1 )^+(Y) \| _{G(\k), h } \leq
 2^{|Y|} \  \sum_{\{Y_j\}}\prod_j \cO(1) f(Y_j)
\ee
and hence
\be  \| (e^{F} -1 )^+ \| _{G(\k), h ,  \G_{p+2}} \leq
\sum_{N=1}^{\infty}(\cO(1))^N \|f\|^N_{\G_{p+4}}
\leq \cO(1)  \|f\|_{\G_{p+4}}   \ee
This gives the result.

\QED
\bigskip

\subsection{Scaling}
In the scaling step we define new polymer activities  $\cS(K)$ so that
\be  \cE xp(\Box + K)(\La,\f_L) = \cE xp(\Box + \cS(K))(L^{-1}\La,\f)
\ee
Here the scaled field is    $\phi_L(x)  =  L^{- \al} \phi(x/L)$ with $\al =
\dim\f =(d-2)/2$.
After a rearrangement one finds
\begin{equation}
  {\cS} (K)(X,\phi) =\sum_{ \{ Y_i \}  \to  LX}
 \prod_i K(Y_i,\phi_L)
\end{equation}
where    the $Y_i$ are disjoint but the $L$-closures
${\bar Y_i^L}$ overlap and fill $LX$.

\bthm Let $c <  (2d\ L^{d/2}\ c_s)^{-1}$  and define $h_L = L^{-\al}h$.
 For any positive $p,q$ there is a constant  $\one$ such
that  \label{scalingthm}
\begin{equation}
  \|\cS(K)\|_{G(\k),h,\Gamma_p} \le \cO (1) L^d \|K\|_{G_L(\k),h_L,\Gamma_{p-q}}
\end{equation}
provided  $\|K\|_{G_L(\k), h_L,\Gamma_{p-q}}$ is sufficiently small.
\ethm

\pr  Let  $Y = \cup_i Y_i $.  Since   $L^{-1}Y \subset X$ we have by a
generalization of  Lemma \ref{disolve} and the bound  $c <  (2d\ L^{d/2}\
c_s)^{-1}$
\begin{equation}  \label{disolve2}
  \prod_i G(\k, L^{-1}Y_i)  = G(\k,L^{-1}Y)   \leq   G(\k,X)
\end{equation}
The point here is that we need the Sobolev inequality on the $L^{-1}$
scale which means that we must replace  $c_s$ by the larger
  $L^{d/2} c_s$.

In the definition of   $\cS (K)$ we write
$K(Y_i, \phi_L)  =  K_{L^{-1}}(L^{-1}Y_i, \phi)  $
and  by (\ref{disolve2}) and  the multiplicative property of the norm we have
\begin{equation}
 \| {\cS} (K)(X)\|_{G(\k),h}   \leq \sum_{ \{ Y_i \}  \to  LX}
 \prod_i  \|  K_{L^{-1}}(L^{-1}Y_i) \| _{G(\k),h}
\end{equation}
However
 $
\|  K_{L^{-1}}(L^{-1}Y) \| _{G(\k),h}
\leq \|  K(Y) \| _{G_{L}(\k),h_L}
$
and so
\begin{equation}
 \| {\cS} (K)(X)\|_{G(\k),h}   \leq \sum_{ \{ Y_i \}  \to L X}
 \prod_i   \|  K(Y_i) \| _{G_{L}(\k),h_L}
\end{equation}

Now multiply by $\G_p(X)$.  By the connectedness we have $\G_p(X)
\leq \prod_i \G_p(L^{-1}{\bar Y_i^L})$.  Furthermore we have the bound
\cite{BrYa90} for some constant $\cO (1)$:
\be\label{gamma2}  \G_p(L^{-1}{\bar Y}^L) \leq \cO (1) \G_{p-q}(Y) \ee
 Summing over $X$
with a pin and using a spanning tree argument we obtain
\be  \|{\cS} (K)\|_{G(\k),h,\G_p} \leq
\sum_{N=1}^{\infty} ( \one L^d\|K\|_{G_L(\k),h_L,\G_{p-q}})^N  \ee
This gives the result.

\QED
\bigskip

\re   The linearization given by
\begin{equation}  \label{s1}
  \cS_1 (K)(X,\phi) =\sum_{ \bar  Y^L  =LX}  K(Y,\phi_L)
\end{equation}
also satisfies the same  bound.

\subsection{Summary}

We combine the three steps into  one  theorem
which tells how the polymer activity changes under a
single RG step.
Our assumptions on the polymer activity   $K$, the extraction $F$, and
parameters  $\kappa , \delta \kappa,  h,  \delta h$ are as follows:
\begin{enumerate}
\item   $ \| K\|_{ G(\k), h , \G}$ is sufficiently small.
\item  The constants  $\kappa,c$ in $G(\k)$ satisfy
  $c \leq (2d\ L^{d/2}\ c_s)^{-1} $ and  $\k c^{-1} L^2$ is
sufficiently small.
 \item  The inequality     $(\de  h)^2 \geq 8\gamma^2\ \|C \|_* \
\|K  \|_ { G(\k), h, \G}$  holds.
\item  The extraction    $F$  is stable for $ (G_{\ell}(\k), h - \de h , f(X))$
and for   $ (G'_{\ell}(\de \k), h - \de h , \de f(X))$
with constants  $f(X), \de f(X)$  such that
$ \|f \|_{\G_{-1}} , \|\de f \|_{\G_{-3}} $ are sufficiently small and such
that  $ \|f \|_{\G_{-1}}  \leq  \one \| K\|_{ G(\k), h , \G}$.
   \end{enumerate}

\bthm   Under the above assumptions  \label{rtheorem}
\be \left(  \mu_C *\cE xp ( \Box + K)( \La) \right) ( \phi_L)
=  \exp \left( \sum_{X \subset \La} F(X, \phi_L)   \right) \
  \cE xp ( \Box + \cR (K,F))(L^{-1}\La, \phi) \label{RGtrans1} \ee
where
\be   \cR(K,F)  =  \cS (\cE (\cF(K),F) ) \ee
 In addition
\be  \|  \cR(K,F)   \|_{ G(\k + \de \k),h - \de h, \G   }
\leq   \one   L^d \|  K   \|_{ G(\k),h, \G   } \ee
\ethm

\pr
If  $K^\# = \cF(K)$ then   by conditions 2,3, Theorem  \ref{flucthm}
is applicable and so
\be\left( \mu_C *  \cE xp ( \Box + K)( \La) \right) ( \phi)
=
  \cE xp ( \Box + K^\#)(\La, \phi)  \ee
and
\be  \label{aaa}
 \| K^\#\|_{ G_{\ell}(\k), h - \de h, \G_{-3}}
 \leq  2 \| K\|_ { G(\k), h, \G} \ee
Then we extract $F$ and we find
\be  \cE xp ( \Box + K^\#)(\La, \phi) =
 \exp \left( \sum_{X \subset \La} F(X, \phi ) \right)
 \cE xp ( \Box + K^*)(\La, \phi) \ee
where   $K^* =  \cE(K^\#,F)  $.  The hypotheses of Theorem
\ref{exthm1} hold for $K^\#$
and
$p=-5$: one has that $\|K^\#\|_{ G_{\ell}(\k), h -\de h,
\G_{-3}}$ is sufficiently small by  assumption 1 and  (\ref{aaa}). Therefore
 \be
\| K^*\|_{ G_{\ell}(\k + \de \k), h - \de h, \G_{-5}}
\leq   \one (  \| K^\#\|_{ G_{\ell}(\k), h - \de h, \G_{-3}}  +\|f\|_{\G_{-1}}  )
\leq   \one \| K\|_ { G(\k), h, \G}
\ee
Finally we scale and find by Theorem \ref{scalingthm} that
\be \cE xp ( \Box + K^*)(\La, \phi_L)  =
 \cE xp ( \Box + K')(L^{-1}\La, \phi)
\ee
where  $K' = \cS (K^*)  = \cR(K,F) $, and since
 $ \| K^*\|_{ G_{\ell}(\k + \de k), h - \de h, \G_{-5}}$ is sufficiently small
we have
\bea   \| K'\|_{ G(\k + \de k), h - \de h, \G}
& \leq & \one L^d \| K^*\|_{ G_L(\k + \de k),( h - \de h)_L, \G_{-5}} \nn \\
& \leq & \one L^d \| K^*\|_{ G_{\ell}(\k + \de k), h - \de h, \G_{-5}} \nn \\
& \leq  & \one   L^d  \| K\|_ { G(\k), h, \G}
\eea
This completes the proof.

\QED
\bigskip

\re The linearization  $\cR_1(K,F) = \cS_1 \cE_1 (\cF_1K,F)     $
 satisfies the same bound.

\section{More estimates}

The last theorem exhibits the obstruction to iterating the RG, namely the
$L^d$ growth factor. The aim in what follows is to
exhibit special cases where one can  beat this growth factor.
There are  three  mechanisms which are more or less model
independent:  higher order terms,
large sets,  and scaling for small sets with extractions.
A fourth mechanism is estimates on the fluctuation integral
for small sets and  charged polymers and is special to the  two
dimensional
sine-Gordon model.  We discuss each of these in turn.

\subsection{Higher order terms}

We show that if  $K,F$  are small enough then  the higher order
terms in
$\cR(K,F)$  are even smaller. This fact, which follows from the next
proposition with
$D=\cO(L^d)$,  will allow us to restrict attention to the
linearized RG. 
\blem
  \label{higherorderlemma}
Suppose that $K,F$ are small enough so that   $sK,sF$
satisfy the hypotheses of Theorem \ref{rtheorem} for all
complex  $s$ in the disc   $|s| \leq D$ for some $D\ge 2$.   Then
\be \cR(K,F)=\cR_1(K,F) +\cR_{\ge 2}(K,F)  \label{expand}\ee
where   $\cR_1(K,F)$ is the linearization and
\be    \label{higherorder}
\|\cR_{\ge 2}( K,F)\|_{G(\k+\de\k),h-\de h,\G}\le \cO(1)
D^{-1}L^d\|K\|_{G(\k),h,\G}
\ee
\elem

\Proof  By Theorem \ref{rtheorem} we have that  $\cR(sK,sF)$ is well-defined
for   $|s|  \leq D $ and satisfies
\be  \label{rs}
\|\cR(sK,sF)\|_{G(\k+\de\k),h-\de h,\G} \leq \cO(1)D L^d\|K\|_{G(\k),h,\G}
\ee
Furthermore it is not difficult to see that
   $ \cR ( sK, sF) $ is analytic
in    $s$ . Expand around $s=0$ and evaluate at $s=1$ and obtain
(\ref{expand})
with the remainder given by
\be\label{cauchy1}
\cR_{\ge 2}(K,F)=\frac{1}{2\pi i}\oint_{|s| = D}\frac{\cR(sK,sF)\ ds}{s^2(s-1)}
\ee
Using  the bound (\ref{rs}) and picking up
an extra factor  $|s^{-2}| =  D^{-2}$ we have the result.

\QED
\bigskip

\subsection{Large sets}
We next study  the linearization $\cR_1(K,F)$ on large sets, that is
on  large  polymers.   A polymer $X$
is called   {\it small} if it is connected and has  $|X| \leq  2^d$.
Otherwise it is a {\it large }  polymer.

The following gives  favourable bounds for  large sets:
\blem   \label{largeset}   Let $K$ be supported on large sets.  Then for
any $p,q >0$
\be   \|\cS _1 (K ) \|_{G,h,\G_p}   \leq
        \cO(1) L^{-2} \|K\|_{G_L,h_L,\G_{p-q}} \ee
Under the hypotheses of theorem  \ref{rtheorem}:
\be \|\cS_1 \cF_1 K \|_{G(\k+\de\k),h-\de h,\G}
\le  \cO(1) L^{-2}\|K\|_{G(\k),h,\G}
\ee
\elem

\Proof    The first bound follows by following  the proof of
 Theorem \ref{scalingthm} for the  linear terms only,  but  replacing
(\ref{gamma2}) by the stronger inequality
\be\label{gammalarge}  \G_p(L^{-1}{\bar X}^L) \leq \cO (1)L^{-d-2}
\G_{p-q}(X) \ee
which is valid for large sets $X$.   This inequality is
proved in \cite{BrYa90} and  \cite{BDH98a}, Lemma 1.

For the second bound we note that if $K$ is supported on
large sets then so is  $\cF_1 K$.   Thus we can use the first
bound followed by our bound on $\cF_1$.
\QED
\bigskip

\re   The second bound gives a good bound on $\cR_1 ( K,F) = \cS_1 \cE_1
(\cF_1 K, F)$
since we will use it in a situation where $\cE_1 (K,F) =K$ and hence
$\cR_1 ( K,F) = \cS_1\cF_1 K$. 
\bigskip
\bigskip
\bigskip

\subsection{Small sets }

For small sets the usual strategy would be to extract  the fastest growing
terms
(the relevant variables) and get good bounds on the remainder.   This generally
works when the canonical scaling dimension of the
field is positive.   However in d=2 the field has dimension
zero and any polynomial in the field is relevant, rendering
the strategy intractable.
For sine-Gordon we  use the fact that  the
interaction  is periodic under translations
$\phi \to \phi + 2\pi $  in field space.
This allows a  Fourier analysis in this translation
variable and a new contraction mechanism for the non-zero Fourier modes.
The remaining zero modes depend only on  $\pa \phi$ which  has a positive
dimension and thus these  terms can be  handled by
extraction. We now give the details.

Let  $K$ be a polymer activity which satisfies
$K(X, \phi + 2 \pi ) =  K(X, \phi  ) $.  Expand
    $  K(X,  \Phi + \phi ) $  in a Fourier series in the
real variable $\Phi$
\be  K(X, \Phi + \phi)= k_0(X,\f)+ \sum_{q\ne 0}  e^{iq\Phi} k_q(X,
\phi)
\ee where
\be \label{kq}
k_q(X,\phi)  = \frac{1}{2\pi}
   \int_{-\pi} ^{\pi} e^{-iq\Phi} K(X, \Phi + \phi) d \Phi .\ee
Then
\be  K(X, \phi)= k_0(X,\f)+ \sum_{q\ne 0}  k_q(X,
\phi).
\label{series}
\ee
The terms with $q\ne 0$ are called the {\em charged} terms and the $q=0$
term is called the {\em  neutral} term. The terminology is consistent
with the Coulomb gas interpretation of the model. We sometimes also
use the notation $\bar K(X,\f)=k_0(X,\f)$.

 Note that for a constant shift $c$
of the field
\be\label{shift}
k_q (X, \phi + c ) = e^{iqc} k_q(X, \phi) .\ee
Also   using $G(\k,X, \phi) = G(\k,X, \phi + \Phi)$  one can show
\be\label{chbound}
\|k_q\|_{G(\k),h,\G} \le  \|K\|_{G(\k),h,\G}.
\ee

\subsubsection{Charged sector}

Now we show how in dimension two only, the charged terms exhibit significantly
improved behaviour under the fluctuation step.

\blem     \label{better}    Let   $K(X, \phi)$ be supported on small sets,
and be  periodic in $\phi$
with Fourier coefficients $k_q(X, \phi)$ as above. Then for $q \neq 0$
\be  \|\mu_C * k_q \|_{G_{\ell}(\k), h, \G_{-1}}
 \leq m_q  \ \| k_q \|_{G(\k), h + N_C, \G}   \label{betterbound}\ee
where
 \bea   N_C &=&  \sup_{X\ \mbox{\rm small}}\  \inf_{x\in X}\|C(\cdot -
x)-C(0)\|_{X}
\nn
\\ m_q  &=&   \exp [ -( |q| - 1/2) C(0) ]. \eea
\elem
\bigskip

\re  The right side of (\ref{betterbound}) can also be bounded by
$m_q\|K\|_{G(\k), h + N_C, \G} $.  Then
if $ k_0  =0$ so that
$ K(X,  \phi) =  \sum_{q \neq 0}   k_q(X, \phi)   $
we have
\be  \|\mu_C * K \|_{G_{\ell}(\k), h, \G_{-1}}   \leq (\sum_{q \neq 0} m_q )\ \| K
\|_{G(\k), h + N_C, \G}
\leq \one e^{-C(0)/2} \| K \|_{G(\k), h + N_C, \G}.
\ee
In $d=2$, $C(0)=\cO(\log L)$, giving a significant decay
factor for $L$ large. In $d>2$, $C(0)=\cO(1)$, and the decay factor
is not significant.
\bigskip

\pr
We have
\be ( \mu_{ C} *  k_q)(X, \phi)=
 \int k_q(X, \phi + \z)
d  \mu_{ C}(\z)  \ee
Now let $f$ be any function  and shift
the integral by
$\z \to  \z  + i \si_q f$
where  $\si_q $ is the sign of $q$.
We find our expression is
\be
e^{ (f,  C^{-1}f)/2}  \nn \\
 \int  e^{-i\si_q (\z,  C^{-1}f)}
 k_q(X,  \phi + \z + i\si_qf) \
d  \mu_{ C}(\z) \ee
Taking  $f(y)=C(y- x)$ where $x$ is an arbitrary point of $X$ gives
\be
e^{  C(0)/2 }
\int    e^{-i\si_q \z(x)} k_q(X,  \phi + \z + i\si_q  C( \cdot - x) )
d  \mu_{ C}(\z) \ee
Now use (\ref{shift})
with  $c = i\si_q C(0)$ to obtain
\be   ( \mu_{ C} *  k_q)(X, \phi)=
m_q
\int    e^{-i\si_q \z(x)} k_{q,x}(X,  \phi + \z )
d  \mu_{ C}(\z)  \label{process} \ee
where
\be  k_{q,x}(X,  \phi  )
= k_{q}(X,  \phi  + i\si_q ( C( \cdot -x)-C(0)) )
\ee
is a translation of  $k_q$.

   Taking derivatives and norms:
  \be
\| ( \mu_{ C} *  k_q)_n(X, \phi)\|   \leq  m_q  \int
\|( k_{q,x})_n(X,  \phi + \z) \|
d  \mu_{ C}(\z)  \ee
 By  Lemma \ref{CX}, $\mu_C * G(\k, X) \leq G_{\ell}(\k , X)2^{|X|}$ and
hence
 \be  \label{thisone}
\| ( \mu_{ C} *  k_q)(X)\|_{G_{\ell},h}   \leq m_q
\| k_{q,x}(X) \|_{G,h} \ 2^{|X|}
 \ee
 (still for any  $x \in X$).
Now in general we can estimate translations by
 \be  \| K (X,\cdot + f) \|_{G,h} \leq   \| K (X) \|_{G,h + \|f\|_{X}} \ee
where  $\|f\|_{X}$ is defined in (\ref{firstnorm}).
This can be seen by making a power series expansion in $f$.
 We apply this to  $k_{q,x}$ and  choose $x \in X$ to minimize  $ \|
C( \cdot
-x)-C(0)\|_{X}$, and find
\be   \label{thatone}
 \| k_{q,x}(X)\|_{G, h}  \leq   \| k_{q}(X)\|_{G, h + N_C}
    \ee
Combining  (\ref{thisone}) and  (\ref{thatone})
gives the result.

\QED
\bigskip

\re   The price we have paid for the strong contraction factor
is a slight loss in
the region of analyticity
$h + N_C \to h$ or $h \to h - N_C$. Iterating this is a problem  in $d=2$
since we do not
recover analyticity in the scaling step. For the UV problem this could be
overcome by taking
$h$ very large  at the start.  However for the IR problem we just have to
do better.

\blem  \label{best}  Let the hypotheses of Lemma \ref{better} hold. For $0
\leq \eta \leq 1$ ,
and any $p,r\ge 0$,
\be  \| \cS_1 k_q \|_{G(\k),h, \G_p}
 \leq \cO(1) L^d e^{ \eta h_L|q|}  \ \| k_q \|_{G_L(\k),h_L(1-\eta/2),
\G_{p-r}}
\ee
\elem
\bigskip

\re   Suppose  $d=2$ so that  $h_L =h$.  The point of the lemma is
that we have traded a slightly worse bound
(the factor $e^{ |q|\eta h}$) for better analyticity (the improvement from
$h(1-\eta/2)$ to $h$).
If we combine  Lemma \ref{better} and
Lemma \ref{best}  with the choice  $\eta=2h^{-1}N_C$ (assumed less than 1)
we find
\bea  \| \cS_1 \cF_1 k_q \|_{G(\k),h, \G}  &=&
 \| \cS_1 (\mu_C * k)_q \|_{G(\k),h, \G}  \nn \\
& \leq &  \cO(1) L^2 e^{ \eta h|q|}
\|  (\mu_C * k)_q \|_{G_L(\k),h(1-\eta/2), \G_{-1}}  \nn \\
& \leq & \cO(1) L^2 e^{ 2N_C |q| }m_q  \| k_q \|_{G(\k),h,
\G}\label{combine}
\eea
Since $C(0) = \cO(\log L)$  and  $N_C = \one$ the factor
 $ m_q  =   \exp ( -( |q| - 1/2) C(0) ) $ is stronger than the factor
 $ e^{ 2N_C |q|}$.  Hence we have accomplished the goal of finding a strong
contraction factor
 without losing analyticity.   (Of course we still have to
see if it is strong enough to beat the factor $L^2$.)
\bigskip

Before embarking on the proof of the lemma we note a
preliminary result which exhibits improved scaling
behavior when a function vanishes at a point.

\blem\label{scalenorm}  Let  $Y$ be a small set in $LX$ and suppose
$f_L(y) =  L^{-\al}f(x/L)$  vanishes at
some point $y_* \in Y$.  Then
\be \| f_{L}\|_Y\le\cO(1)L^{-1- \al }\| f\|_{X}  \ee
\elem

\nind \pr   First observe that $\pa^\beta
(f_{L}(x))=L^{-|\beta|-\al}
(\pa^\beta f)(L^{-1}x)$, so we need only look at the nonderivative term in
the norm.
Now note that  for any
$y\in Y$ the length of the shortest rectilinear path
within $Y$ from $y$ to
$y_*$, is less than $\cO(1)$. Therefore  since  $f_L$ vanishes at $y_*$
\bea \ | f_L(y)|&   =& | f_L(y)  -  f_L(y_*)|
\nn \\
& \leq &  \one
\sup_{z \in Y,|\beta|=1}|\pa^\beta f_L(z)| \
\nn\\ &\le&{\cO(1)}L^{-1-\al}\
\| f\|_{X} \eea
\QED

\nind{\bf Proof }(of Lemma \ref{best}):  With $k_q' =   \cS_1  k_q$   we have
\be \label{kprime}  k_q'(X,\phi)  = \sum_{Y:\bar Y^L=LX}   k_q(Y,  \phi_{L})
\ee
where the sum is over small sets.
For each term of (\ref{kprime}) we use   (\ref{shift}) to  shift $\f_L$ by a
constant
$\eta\f_L(y_*)$ where $y_*$ is an arbitrary point of $Y$. Then we have
\be    k_q'(X, \phi)=  \sum_{\bar Y^L=LX}
e^{iq \eta \phi_L(y_*) }
 k_{q}(Y, (1-\eta) \phi_L + \eta \tilde\phi_L ) \ee
 Here we have defined   $\tilde f(x) = f(x) - f(y^*/L)$ so that
$\tilde f_L(y)  = f_L(y) - f_L(y^*) $.
 Lemma
\ref{scalenorm} implies
\be  \|(1-\eta) f_L + \eta  \tilde f_L\|_Y  \leq L^{-\al }[ (1-\eta)  +
(\cO(1)/L) \eta] \leq L^{-\al}[1-\eta/2]\ee
whenever  $\|f \|_{X} \leq
1$ and so when computing derivatives we obtain
\be   \|( k_q')_n(X, \phi)\|   \leq   \sum_{\bar Y^L=LX} \sum_{a+b=n}
\frac{n!}{a!b!}
L^{-n\al}(|q|\eta )^a (1-\eta/2)^b \|( k_{q})_b(Y,(1-\eta) \phi_L +
\eta \tilde \phi_L  )\|\ee
We also have by  (\ref{disolve2})
\be  G_L(\k,Y,(1-\eta) \phi_L + \eta \tilde \phi_L ) = G(\k,L^{-1}Y, \phi)
\leq G(\k, X, \phi) \ee
and so
\be   \|( k_q')_n(X)\|_{G} \leq    \sum_{\bar Y^L=LX} \sum_{a+b=n}
\frac{n!}{a!b!}
L^{-n \al }(|q|\eta)^a (1-\eta/2)^b  \|( k_{q})_b(Y )\|_{G_L}
 \ee
and so
\be \| k_q'(X)\|_{G, h} \leq  e^{\eta h_L|q|}    \sum_{\bar Y^L=LX}
  \| k_{q}(Y)\|_{G_L, h_L(1-\eta/2)}
 \ee
The rest of the proof follows as in theorem \ref{scalingthm}.

\QED

\subsubsection{Neutral sector}

Improved bounds can be arranged for general activities defined on small sets by
extracting a finite number of terms characterised by low ``scaling
dimension''.   As in
\cite{BDH98a} we define
 the  {\it scaling dimension} $\dim K$  of any polymer activity $K$
by
\bea
        \dim ( K_{n} )
&=&
        r_n + n \dim \f;
\nn \\
        \dim  (K)
&=&
        \inf_{n} \dim ( K_{n})
\label{dimK}
\eea
where the infimum
is taken over $n$ such that $K_n(X,0)  \neq 0$. Here $r_n$ is defined to
be  the
largest integer satisfying $r_n \leq r$ and
$K_{n}(X,\phi=0;p^{\times n})=0$ whenever
$p^{\times n}=(p_1,\dots,p_n)$ is an $n$--tuple of polynomials of total
degree less than $r_n$.  One can interpret $r_n$ as the
number of derivatives present in the  $\phi^n$ part of $K$  (up to a
maximum  $r$).

For comparison purposes we quote the following result from
\cite{BDH98a}:

\bthm\label{improved}  Suppose $d \geq 3$,, $K(X,\phi)$ is supported on small
sets, and  $\k h^2 \geq \one$.   Then  for any $p,q\ge 0$ there is a constant
$\one $
such that
\be
        \|\cS _1 (K) \|_{G,h,\G_p}
\leq         \one L^{d-\dim (K)} \|K\|_{G_L,h,\G_{p-q}}
\ee
\ethm

The proof needs  $\dim \phi >0$ and fails for  $d=2$.  However we can obtain
a similar result for  $d=2$ if we restrict to the neutral sector.

\blem \label{good}    Suppose $d=2$,   $K(X,\phi)$ is supported on small
sets
and satisfies the neutrality condition $K(X,\phi+c) =  K (X,
\phi)$ for
any real c, and that
$\kappa h^2 \geq \one $.
  Then  for any $p,q\ge 0$ there is a constant  $\one $
such that
\be  \| \cS_1 (K) \|_{G,h, \G_p}
\leq  \one L^{2-  \dim (K)}   \| K \|_{G_L,h, \G_{p-q}}   \ee
\label{specialone}
\elem

\re   The neutrality condition implies $K_n(X, \phi;f_1,...,f_n)$
vanishes if
any $f_i$ is a constant.  Hence
$ \dim K_n =r_n \geq n $ for  $n <r$ and  $ \dim K_n =r_n=r $ for  $n \geq r$.
   \bigskip

\pr  Starting from the definition (\ref{s1})
we have
\be   (\cS_1K)_n(X,\phi)  = \sum_{Y:\bar Y^L=LX}(K_{L^{-1}})_n(L^{-1}Y,
\phi)
\ee
Thus we need to estimate
\be  \| (K_{L^{-1}})_n(L^{-1}Y,\phi)  \|  = \sup_{\|f_i \|_{X} \leq 1}
 |K_n(Y, \phi_{L};
 f_{1,L}, ... , f_{n,L})|
\ee

By the remark above  the supremum can be taken over fields $f_i$ such that
$f_{i,L}$ vanishes at a point in $Y$. For such fields Lemma
\ref{scalenorm}
applies again giving $\|f_{i,L}\|_{Y}\le \one L^{-1}\|f_i\|_{X} $
and it follows that
\be
 \|(\cS_1K)_n(X,\phi) \| \leq  \sum_{Y}
\| K_n(Y, \phi_{L}) \| (\one L^{-1}) ^n
  \ee
We proceed as in the proof of Theorem \ref{scalingthm},
first summing only over $n \geq \dim (K)$ so we can gain
a factor $L^{-\dim(K)}$.  With  $\dim (K) =k $ we  have
\be  \sum_{n \geq k} h^n/n!
 \|(\cS_1K)_n(X) \|_{G} \leq  \one   L^{-k}\sum_Y
\| K(Y) \|_{{G_L},h}   \label{high}
  \ee

We  do something different for derivatives $K_n$ with $n <k$.
We have
the representation
\bea\label{special}
K_n(Y,\f_L;f_{L}^{\times n})
&=&
\sum_{m=n}^{k-1}\frac{1}{(m-n)!}  K_m(Y,0;  \phi_L^{\times (m-n)}\times
f_L^{\times n}  )\nn \\
&+&
\int^1_0\ dt\
\frac{(1-t)^{k-n-1}}{(k-n-1)!}\
   K_{k}(Y,t\f_{L}; \f_{L}^{\times (k-n)}\times
f_{L}^{\times n})  \eea   Again by the neutrality condition we can replace
$\phi_L$ by  $\tilde \phi_L(y)
= \phi_L(y) - \phi_L(y_*) $ for some $y_* \in Y$,  and similarly for   $f_L$.
Now in \cite{BDH98a}, Lemma 15, it is proved that
\be   | K_n(Y,0;  f_L^{\times n}  ) |  \leq  (\one)^n  L^{-\dim K_n}
 \| K_n(Y,0 ) \| \ \prod_{j=1}^n \|f_j\|_{X} \ee
Use this bound on the terms in the sum. The  remainder is estimated
using   $\| \tilde \phi_L \|_{Y}  \leq \one L^{-1} \| \tilde \phi \|_{X} $
from Lemma \ref{scalenorm}.
We find
\bea
&& \| (\cS_1 K)_n(X,\f) \|  \leq
\one L^{-k}\sum_{Y} \nn \\
&&  \left\{    \sum_{m=n}^{k-1}
  \| K_m(Y,0)\| \  \| \tilde  \phi \|_{X}^{m-n}
+
\int^1_0\ dt\
(1-t)^{k-n-1}\  \| K_k(Y,t\f_{L})\|\   \| \tilde  \phi \|_{X}^{k-n}
\right\} \eea
Now multiply by  $G(\k, X, \phi)^{-1} $.  For the remainder term we use
\bea  G(\k,X, \phi)^{-1}
&=&  G(\kappa  t^2, X, \phi)^{-1}   G(\kappa (1- t^2), X, \phi)^{-1}  \nn \\
& \leq &  G_L(\kappa  t^2, Y, \phi_L)^{-1}   G(\kappa (1- t^2), X, \phi)^{-1}
\eea
where we have used
(\ref{disolve2})  again.  We next use
\be   \sup_\f \ \| \tilde  \phi  \|^{a}_{X}\ G(\kappa (1- t^2), X, \phi)^{-1}
 \leq   \one (\kappa(1-t^2))^{-a/2}  \ee
 This is a Sobolev inequality on derivatives of order up to
$r$ and needs   $s >d/2 + r$.   For the zeroeth derivative we  dominate
$ \tilde \phi$ by a first derivative  and  then use the Sobolev inequality.
Here we use the fact that $X$ is necessarily small and so has diameter  $\one$.
Now the integral over  $t$ can be estimated by
 $ \one \| K_k(Y) \|_{G_L}  \k ^{-(k-n)/2}$.   The terms in the sum over
$m$ are treated
similarly and we end up with   \be  \| (\cS_1K)_n(X) \|_G
\leq  \one  L^{-k} \sum_{Y}  \sum_{m=n}^k
  \| K_m(Y)\|_{G_L}  \  \kappa^{-(m-n)/2}     \ee
 Since  $\kappa^{-1/2}  \leq \one h$, this leads for  $n<k$ to
\be \frac{h^n}{n!} \| (\cS_1K)_n(X) \|_G \nn \\
\leq  \one  L^{-k} \sum_{Y}
  \| K(Y)\|_{G_L,h}     \ee
Combining this with (\ref{high}) we find
\be  \| (\cS_1K)(X) \|_{G,h}
\leq  \one  L^{-k} \sum_{Y}
  \| K(Y)\|_{G_L,h}     \ee
and the result follows as before.

\QED \bigskip

\section{The infrared problem}
We return to the sine-Gordon model in $d=2$.
The infrared problem for
$\beta > 8 \pi$ is to study the partition
function
\begin{equation}  \label{Z6}
Z=\int \exp\left(\z\int_{\La_M} \cos(\f(x))dx\right)\ d\mu_{\beta v^M_{0}}(\f).
\end{equation}
in the limit $M \to \infty$.  In particular we  want to
prove Theorem \ref{irthm1}.

We shall use a family of polymer activity norms defined for $j=0,1,2,...$ by
\be \|K\|_j=\|\cdot\|_{G(\k_j),h_j, \G}  \ee
 where the underlying $\f$--norms in (\ref{firstnorm}) are taken with $r=4,s=6$.
The large field regulator is
$G(\k_j)$  defined by  (\ref{gee}) with
\be \kappa_j=\kappa_0 \left(\sum^{j}_{k=0} 2^{-k}\right)
\ee
 We choose  $c=(8 L c_s)^{-1}$ and $\k_0c^{-1} L^2$  sufficiently small that
Lemma \ref{CX} holds (thus
$\k_0
\leq
\cO(L^{-3}$)). Note that $\k_j$ increases slowly in $j$.
The domain of analyticity is defined by
\be h_j=h_{\infty} \left(1+\sum^{\infty}_{k=j+1} 2^{-k}\right)
\ee
with  $h_{\infty}=\k_0^{-1/2}$ (so  $h_{\infty} \geq \cO (L^{3/2} )$).
Note that  $h_j$ decreases slowly in $j$ and that
$\k_j^{1/2}h_{j'} \geq \k_0^{1/2}h_{\infty} =1$.
Finally $\G$ is defined as  in (\ref{gamma}).
We restate Theorem  \ref{irthm1} as follows:

\bthm\label{irthm2} Let $\beta $ be chosen from a compact subset of
$(8\pi,\infty)$, let $0<\ep<1$, and let $L$ be chosen sufficiently large.
Then
there is a number $\bar\z$ such that for all
$\z$ real with $|\z|\le \bar\z$ and any $0\le j\le M$ the
partition function has the form
\be Z=e^{\cE_j}  \int \cE xp ( \Box  + K_j)(\La_{M-j},\f)
 d \mu_{\beta v^{M-j}_0(\si_j)}(\phi) \label{Z}
\ee
where the polymer activities $K_j$ are  translation invariant, and
 even and $2\pi$--periodic in $\f$.
They satisfy the bounds
\be \|K_j\|_j \le
 \de^j|\z|^{1-\ep}
\ee
where $\de= \one \max\{L^{-2},L^{2-\beta/4\pi}\}<1/4$.  Furthermore
the energy density and the field strength have the form
\bea   \cE_j &=&  \sum_{k=0}^{j-1}     \de \cE_k    \nn \\
  \si_j &=&  \sum_{k=0}^{j-1}  \de \si_k  \eea
and satisfy the bounds
 \bea
|\de \cE_k | &\leq& \one \    \de^k\ |\z|^{1-\ep} |\La_{M-k}| \nn \\
|\de \si_{k}| &\leq& \one\  h_{\infty}^{-2}\   \de^k\ |\z|^{1-\ep}
\label{rbd}
\eea
\ethm
\bigskip

\re Since
$\|K_j\|_\infty\le
\|K_j\|_j$ the version stated in  Theorem  \ref{irthm1} follows as well.
\bigskip

\pr  The proof is by induction on $j$.  For  $j=0$ we  write the
interaction as a sum over unit blocks,  make a Mayer expansion,
and then group together into connected components to obtain
\be   \exp ( \sum_{\De \subset \La_M} \z V(\De)  )
= \sum_{ \{ \De_i\} }  \prod_{i}  (e^{\z V(\De_i)} -1)
=  \cE xp ( \Box + K_0 )  ( \La_M)
\ee
Here $K_0$ is supported on connected polymers and is given by
\be   K_0 (X)  = \prod_{\De \subset X} (e^{\z V(\De)} -1)\ee
 However
by Lemma \ref{vbd} in the appendix
we have the estimate for  $|\z| $ sufficiently small
 \be   \|  e^{ \z V(\De)}  -1 \|_{1,h_0}   \leq  |\z|^{1-\ep/2} \ee
 Hence  $ \| K_0(X) \|_{1,h_0}     \leq  (|\z|^{1-\ep/2} )^{|X|} $
and it  follows by a standard bound \cite{DiHu91} that    $\| K_0\|_0  \leq
 |\z|^{1-\ep}  $.
 Thus the representation
and the bound hold for $j=0$.

Before proceeding to the general step of the induction we
specify the extractions we want to make.
 For an expression $\cE xp (\Box + K)(\La,
\phi)$, the extracted part $F=F(K)$ is taken from the
 neutral sector $ \bar K(X, \phi)  =(2\pi)^{-1}
   \int_{-\pi} ^{\pi} K(X, \Phi + \phi) d \Phi  $ for small sets.
It is chosen satisfying $F(X, \phi + c) = F(X, \phi)$
and so that  $\dim(\bar K -F)$ is
larger than zero. In fact we want to  choose $F$ so that    $\dim(\bar K -F) \geq
4$ ( this is why we need $r=4$). These conditions are more than sufficient to beat
the factor
$L^2$ in the scaling step.
As noted earlier the neutrality condition  implies
$\dim(\bar
K_n)\ge \min ( n, 4)$ ,  and hence we may take
$F_n=0$ for $n  \geq 4$. Also note that $\bar K_n(X,0)=0$ for $n$ odd,
and  hence we may take
$F_1,F_3=0$.
The  remaining
conditions are for small sets  $X$:
\begin{eqnarray}  \label{cond}
(\bar K - F)_0(X,0)&=&0  \nn \\
(\bar K - F)_2(X,0; x_{\mu} , x_{\nu})&=&0  \nn \\
(\bar K - F)_2(X,0;x_{\mu}, x_{\nu} x_{\rho})&=&0
\end{eqnarray}
If we define
the extracted part by   $F(X) = \sum_{\De} F(X, \De)  $ and
\be F(X, \De,
\phi)=\alpha^{(0)}(X)+\sum_{\mu,\nu}\alpha^{(2)}_{\mu,\nu}(X)
\int_{\De}(\pa_\mu \f)(\pa_\nu\f)+
\sum_{\mu,\nu\r}\
\alpha^{(2)}_{\mu,\nu\r}(X)\int_{\De}(\pa_\mu\f)(\pa^2_{\nu\r}
\f) \label{F}
\ee
then the conditions (\ref{cond}) determine
\begin{eqnarray}   \label{alphas}
\alpha^{(0)}(X)&=&|X|^{-1}\ \bar K_0(X,0) \ 1_{\cS}(X) \nn \\
\alpha^{(2)}_{\mu,\nu}(X)&=&(2|X|)^{-1}\bar K
_2(X,0;x_\mu,x_\nu)     \ 1_{\cS}(X) \nn \\
\alpha^{(2)}_{\mu,\nu\r}(X)&=&|X|^{-1}\bar K_2(X,0;x_\mu,x_\nu x_\r) \
1_{\cS}(X)
\end{eqnarray}
where  $1_{\cS}$ is the characteristic function of small sets.
The last two equations define  $F=F(K)$.
\bigskip

Now we continue with the induction, supposing the theorem is true for $j$
and proving it for $j+1$.
The RG applied to
 $\cE xp ( \Box  + K_j)(\La_{M-j},\f)$ starts with a fluctuation integral
with the
measure   $\mu_{\beta C_j}$ where
\be   C_j(x-y)  =  v_0^{M-j} ( \si_j, x-y) -  v_0^{M-j-1} ( \si_j, (x-y)/L)
\ee
 Let  $\cF_j$ be the map on polymer
activities associated with this operation, so  the new activities are
$ K_j^\# =\cF_j(K_j)$.
 Next we extract
$F_j = F(K_j^\# )$ with coefficients  $\al_j$ as specified above. Finally we scale
to the volume
$\La_{M-j-1}$.   Thus  as in Theorem
\ref{rtheorem}:
\bea && \biggl (  \mu_{\beta C_j} *\cE xp(\Box + K_j)(\La_{M-j})\biggr)(\f_L)
\nn
\\ & &= \exp \left( \sum_{X \subset \La_{M-j}} F_j(X,\f_L)  \right)
\cE xp(\Box + K_{j+1})(\La_{M-j-1}, \f ) \label{what} \eea
where
\begin{equation}
K_{j+1}  =  \cR_j(K_j)  \equiv  \cS (\cE (K_j^\#,F(K_j^\#  ))
\end{equation}

Using the  lattice invariances one can prove that
 \begin{eqnarray}
\sum_{X\supset \De}\alpha^{(0)}_{j}(X)&=&\de E_j  \nn \\
\sum_{X\supset \De}\alpha^{(2)}_{j,\mu,\nu}(X)&=&
-(2\beta)^{-1}\ \de_{\mu\nu}\ \de\si_j
\nn \\
\sum_{X\supset \De}\alpha^{(2)}_{j,\mu,\nu\rho}(X)&=&0
\end{eqnarray}
for  some  constants $\de E_j,\de\si_j$.
Now  (\ref{what}) becomes  \bea && (  \mu_{\beta C_j} *\cE xp(\Box +
K_j)(\La_{M-j}))( \f_L) \\
& =& \exp \left(   \de E_j
|\La_{M-j}| -  \frac {\de \si_j}{2 \beta}
 \int_{\La_{M-j-1}} (\pa \phi)^2  \right)
\cE xp(\Box + K_{j+1})(\La_{M-j-1}, \f )  \eea
The partition function $Z$ is the integral of this with respect
to  $ \mu_{\beta v^{M-j-1}_0(\si_{j})}$.  Absorbing the $\int (\pa \phi)^2$
term into this measure changes $v(\si_j)$ to $v(\si_{j+1})$ with
\be    \si_{j+1}   =  \si_j + \de \si_j   \ee
and we have
\be Z =e^{\cE_{j+1}} \int
 \cE xp ( \Box + K_{j+1})(\La_{M-j-1},\phi) d \mu_{\beta
v^{M-j-1}_0(\si_{j+1})}(\phi)
\ee
where \be     \label{de} \cE_{j+1}=\cE_{j}+\de E_j|\La_{M-j}|
+ \log \left[
 \int \exp \left(
 \frac {-\de \si_j}{2 \beta} \int_{\La_{M-j-1}} (\pa \phi)^2  \right) d
\mu_{\beta
v^{M-j-1}_0(\si_{j})}(\phi)   \right]
\ee
This establishes the required form (\ref{Z}) for $j+1$.

Theorem \ref{rtheorem} will be used to obtain a
crude bound on $\|K_{j+1}\|_{j+1}$.  With $\de h_j = h_j-h_{j+1} =
2^{-j-1}h_{\infty}$
and  $ \de \k_j =  \k_{j+1} - \k_j= 2^{-j-1}\k_0$  we check the hypotheses
of this theorem.
\begin{enumerate}
\item
 This is true by the inductive assumption on $K_j$ for $\bar\z$
sufficiently
small.
\item
 True by our choice of $\kappa_0, c$.
\item
 First note from Lemma
\ref{covariancebound} in the appendix that $\|\beta C_j\|_*$ is bounded
 uniformly in $j$.
 Also $\de^j(\de h_j)^{-2}$ is bounded uniformly in $j$ for $L$ sufficiently
large, and therefore \be
\|K_j\|_j \le \de^j|\z|^{1-\ep} \le (8\g^2\|\beta C_j\|_{*})^{-1}
(\de h_j)^2
\ee
holds for all $j$ provided $\bar\z$ is small enough.
\item

The stability conditions will be verified by using Lemma \ref{fbound} in the
appendix which involves \be \|\alpha(X)\|_a= |\al^{(0)}(X)|
 + a^2 \sum_{\mu \nu} |\al^{(2)}_{\mu\nu}(X)| +   a^2
\sum_{\mu \nu \rho}| (\al^{(2)}_{\mu\nu \rho}(X)|
\ee
By this lemma  $F_j$ is stable for     $(G_{\ell}'(\k_j), h_{j+1},
f_j(X))$ if we take the definition $f_j(X)= \one\|\al_j(X)\|_{h_{j+1}}$.  We
need 
$\|f_j
\|_{\G_{-1}}$  small and   $\|f_j \|_{\G_{-1}} \leq \one
\|K_j \|_j $ and it
suffices to show the latter.
Now in the definition  of  $\alpha_j(X)$ replace  $x$ by  $x-x_*$
where $x_*$ is some point in $X$.  Then we obtain the estimates for $n=0,2$
\be  \label{alphabound} | \al_j^{(n)}(X)|  \leq  \one \| \bar K^{\#}_n(X,0)
\|
 \leq  \one \|  K^{\#}_n(X,0) \| \leq
  \one \|  K^{\#}_n(X) \|_{G_{\ell}(\k_j)} \ee
It follows that
\be    \| \al_j(X)\|_{h_{j+1}}  \leq   \one  \| K^{\#} (X) \|_ {G_{\ell}(\k_j),
h_{j+1}}\label{alphabound2}  \ee
and hence \be  \| f_j\|_{\G_{-3}}  \leq   \one  \| K^{\#} \|_ {G_{\ell}(\k_j),
h_{j+1},
\G_{-3}}
\leq   \one  \| K_j  \|_j \ee
Since  $f$ is supported on small sets the same bound holds for  $ \|
f_j\|_{\G_{-1}}$.

Lemma \ref{fbound}  also says that  $F$ is stable for
  $(G_{\ell}( \de \k_j), h_{j+1}, \de  f_j(X))$
if we define   $ \de f_j(X)=  \one  \|\al_j(X)\|_{ \de \k_j^{-1/2}}$.   We must
show that
$\|
\de f_j \|_{\G_ {-3}} $ is sufficiently small under
our hypotheses. We have that
 $1
\leq
 \de \k_j^{-1} h_{j+1}^{-2}
\leq 2^{j+1} $
and hence  $|\de f_j(X) | \leq  \one 2^j | f_{j}(X)| $. Therefore
\be |\de f_j |_{\G_{-3}} \leq \one 2^j | f_j |_{\G_{-3}}
\leq \one 2^j  \|K_j\|_j\leq   \one (2\de)^j | \z |^{1-\ep} \ee which is
small for $\z$ small. \end {enumerate}

This verifies the hypotheses of Theorem \ref{rtheorem} and we
conclude
\begin{equation}
\| K_{j+1}\|_{j+1}  =\| R_j(K_j)\|_{j+1}  \leq   \one L^2 \|K_j\|_j
\end{equation}

It remains to improve the crude bound on  $K_{j+1}$ to
$\| K_{j+1}\|_{j+1}\le \de \|K_j\|_j$ so we get the required
$\|K_{j+1}\|_{j+1}\le \de^{j+1}|\z|^{1-\ep}$.
To accomplish this
let   $1_{\cS}$ (respectively $1_{ \bar \cS}$) be the characteristic
function of small  (large) sets,  write   $K_j =  \sum_q k_q $ as in
(\ref{series}), and make the decomposition
\be\label{4terms}
K_{j+1}=\cR_{\ge 2}(K_j) + \cR_1( K_j 1_{\bar \cS})
+ \cR_1(\sum_{q\ne 0} k_q1_\cS)  +  \cR_1(k_0 1_\cS)
\ee
   We will
show
that each of the four terms on the right can be bounded by
$ (\de/4)  \|K_j\|_j  $.

\begin{enumerate}
\item As above one can check that
Theorem  \ref{rtheorem}  holds for  $sK_j, sF_j$ with  $|s| \leq
L^4$. Then by
 Lemma \ref{higherorderlemma} with $D= L^4$
\be
\|\cR_{\ge 2}(K_j)\|_{j+1}\le \cO(1) L^{-2}\|K_j\|_j \leq  \frac\de 4\|K_j\|_j
\ee

\item The extraction is zero on large sets and so by Lemma \ref{largeset}
\be
\|\cR_1( K_j 1_{ \bar \cS})\|_{j+1}
= \|\cS_1 \cF_1 ( K_j 1_{ \bar \cS})\|_{j+1}
\le\one L^{-2}\|K_j\|_j\le \frac\de 4\|K_j\|_j
\ee

\item
There is no extraction in  $\cR_1 (k_q 1_{\cS})$ since the
extraction is based on  $\overline{ \cF_1 (k_q1_{\cS} ) }
 =  \cF_1 ( \bar k_q 1_{\cS}) = 0$. Hence the third
term is
$\sum_{q\ne 0} \cS_1\cF_1 ( k_q\ 1_\cS)$ which we bound by putting together
Lemmas
\ref{better},
\ref{best}.  As in (\ref{combine}) we have
\be   \| \cS_1 \cF_1 ( k_q 1_{\cS}) \|_{j+1}
\leq  \cO(1) L^2 e^{ 2N_{\beta  C_j} |q| }
e^{-(|q| - 1/2)\beta C_j(0)}  \| k_q \|_j
\ee
However by estimates on $C_j$ in  Lemma \ref{covariancebound} in the
Appendix we have
\be N_{\beta C_j} \le \beta \|\pa C_j\|_\infty \le
\one  \ee
and \be
C_j(0)=  \frac{ \log L}{2\pi(1+\si_j)} + \cO(e^{-L^{M-j-1}/2}).\ee
Using also  $ \| k_q \|_j \leq  \|K_j\|_j$  and the bound on $\si_j$
we have for
 $L$ sufficiently large:
\bea
\|\cR_1  \left(\sum_{q\ne 0} k_q1_\cS  \right)\|_{j+1}
&\le & \one L^2 \sum_{q\ne 0}
\biggl(e^{-|q|(\beta C_j(0)-2N_{\beta C_j})+\beta C_j(0)/2}\biggr)\
\|K_j\|_j  \nn\\
& \leq & \one L^{2-\beta/4\pi}\|K_j\|_j  \nn\\
& \leq & \frac\de 4\|K_j\|_j\eea

\item This term has the desired bound because of the extraction.
Let  $K^{\dagger} =  \cF_1(k_0 1_{\cS})$.
Then we have
  $\cR_1 (k_0 1_{\cS})  = \cS_1 ( K^{\dagger}  - F( K^{\dagger} )) $.
The extraction $F$ is defined so that
$\dim(\bar  K^{\dagger}  - F(  K^{\dagger} ) ) \geq 4$,
but we have   $\bar K^{\dagger} = K^{\dagger}$  (since the same is 
true of $k_0$) and hence  $\dim( K^{\dagger}  - F(  K^{\dagger} ) ) \geq 4$.
Then Lemma
 \ref{good} applies (note $\k_{j+1}h^2_{j+1} \geq 1$)
and gives
\be
\|\cR_1(k_0 1_\cS)\|_{j+1}
\leq    \one L^{-2}
\|  K^{\dagger}  - F( K^{\dagger}) \|_{ G_{\ell}(\k_{j+1}), h_{j+1}, \G_{-3}}
\ee
Now   $\|  K^{\dagger} \|_{ G_{\ell}(\k_{j+1}), h_{j+1},
\G_{-3}} \leq \one \| K_j \|_j  $ .  Furthermore the same bound 
holds for $ F(K^{\dagger}) $.  To see this
extend  the
argument of Lemma \ref{fbound}  in the appendix.
If 
$\alpha ^{\dagger}$ is defined from $K^{\dagger}$ we argue as in  
(\ref{Fbound}) and (\ref{alphabound2}) and  find
\be \|(F(K^{\dagger}))(X)\|_{G_{\ell}(\k_{j+1}),h_{j+1}} \le \one\
\|\al^{\dagger}(X)\|_{h_{j+1}}  \leq  \one  \|K^{\dagger}(X)
\|_{G_{\ell}(\k_{j+1}),h_{j+1}}
\ee which is enough. Thus
\be \|\cR_1(k_01_\cS)\|_{j+1}\le
 \one L^{-2}\|K_j\|_j  \le  \frac\de4\|K_j\|_j
\ee
\end{enumerate}

This  completes the bound on   $\| K_{j+1} \|_{j+1}$.
The last step is to establish   the bounds
(\ref{rbd}).  Using (\ref{alphabound}) we have
\bea
|\de E_j|&\le& \one \| K^{\#}_0\|_{G_{\ell}(\k_j) ,\G_{-3}}
\leq  \one \|K_j\|_j  \leq  \one  \de^j | \z |^{1- \ep}
 \nn\\
|\de \si_j|&\le& \one \beta\| K^{\#}_2\|_{G_{\ell}(\k_j) ,\G_{-3}}
\leq  \one h_{j+1}^{-2}\|K_j\|_j  \leq  \one h_{\infty}^{-2} \de^j | \z
|^{1- \ep}
\eea
We also need to bound $\de\cE_j$. 
Let $v=v^{M-j-1}_0(\si_j)$ and let   $T= v^{1/2}\De v^{1/2}$,
a positive self-adjoint operator.
Doing the integral  in   (\ref{de}) we find 
 \bea  \de \cE_j &=& \de E_j |\La_{M-j}| 
 + \log \left( \det(1 + \de \si_j T)^{-1/2}\right) \nn \\
&=&  \de E_j |\La_{M-j}| 
 - \frac12  tr \left(  \log(1 + \de \si_j T) \right)
 \eea 
But  $\|T\| \leq 2$ and $|\de \si_j |$ is small so the 
spectrum of   $\de \si_j T$ is confined to a 
small neighborhood of the origin.  Hence  $| \log ( 1 + \la)|
\leq  \one |\la| $ for any eigenvalue $\la$ and hence 
\be  | tr \left(  \log(1 + \de \si_j T) \right)| \leq 
\one  tr( |\de \si_j T| ) = \one |\de \si_j | tr(T ) 
\leq   \one |\de \si_j | |\La^{M-j-1}|  \ee
where the last step is an explicit computation.
 Now the bounds  on  $\de E_j$ and $\de \si_j$ yield the bound
$ |\de \cE_j | \leq  \one  \de^j | \z |^{1- \ep}|\La_{M-j}|$.
This completes the proof of the infrared theorem.

\QED
\bigskip

\section{The ultraviolet problem}
The ultraviolet problem on the unit torus
$\La_0$ for  $\beta < 8 \pi$ is equivalent to a scaling limit for unit cutoff
theories.  Thus  we study the $N \to \infty$
limit of the partition function
\be Z = \int  \exp  \left( \z_{-N} \int_{\La_N}  \cos (\phi(x) ) dx  \right)
 d \mu_{\beta v^N_0}(\phi) .\ee

After a number of RG transformations the RG index will increase from $-N$ to a
value $j\le 0$ and  we will be on a volume
$\La_{|j|}$ with a  coupling constant which will have grown  from the ultra small
$\z_{-N}$ to 
\be
  \z_{j } =
 L^{-2|j|} e^{\beta v^{|j|}_0(0)/2}\z .
\ee
At this point
polymer activities are estimated with a norm essentially the
same as for the IR problem, but with relaxed smoothness in $\f$
characterized by $r=2,s=4$ in (\ref{firstnorm}).  We take
\be \|\cdot\|_j=\|\cdot\|_{G(\k),h_{j},\G}  \ee
with  $c=(8Lc_s)^{-1}$, $\k=\cO(L^{-3})$ sufficiently small
so that Lemma \ref{CX}
holds, and 
\be h_j=h_0\biggl[1+\sum_{k=1}^{|j|} 2^{-k}\biggr] \ee
(which decreases in j) 
with  $h_0=\k^{-1/2}=\cO(L^{3/2})$, and $\G$ as in (\ref{gamma}).

Our aim is now to prove the following refinement of Theorem \ref{uvthm}.
\bthm\label{uvthm2} Let $\beta $ be chosen from a compact subset of
$(0,8\pi)$, let $0<\ep<1/4$, and let $L$ be chosen sufficiently large. Then
there is a number $\bar\z$ such that for all
$\z$ complex with $|\z|\le \bar\z$ and any $-N\le j\le 0$ the
partition function has the form
\be Z=e^{\cE_j}  \int \cE xp ( \Box  + K_j)(\La_{|j|},\f)
 d \mu_{\beta v^{|j|}_0}(\f) \ee
 The polymer activities $K_j$ are translation invariant, even and
$2\pi$--periodic
 in $\f$, analytic in $\z$ and have the form
\be K_j=\z_j V +\tilde K_j \ee
 where  $V$ is given by (\ref{vdefn}).  We have the estimates
\bea \| \z_j  V \|_{j}  &\leq & |\z_j |^{1-\ep} \nn \\
\|\tilde K_j\|_j &\le& |\z_j|^{2-4\ep} \eea
Furthermore, the energy density  has the form \be \cE_j =
\sum_{k=-N}^{j-1} \de E_k  |\La_{|k|}|  \ee
 where
\be |\de E_k | \leq \one |\z_k|^{2-4\ep}  \label{debd} \ee
\ethm

\bigskip

\Proof
 The proof is by induction on $j= -N, ...-1$.  For
$j=-N$  the initial density can be written just as in the IR problem as
 \be   \exp (\z_{-N} \int_{\La_N} \cos \phi  )
=  \cE xp ( \Box + K_{-N} )  ( \La_N,\f)
\ee
where $K_{-N}$ is supported on connected polymers and given by
\be   K_{-N} (X)  = \prod_{\De \subset X} (e^{\z_{-N} V(\De)} -1)\ee
If   $X= \De$  we write  $  K_{-N}(\De)=\z_{-N} V(\De) +\tilde K_{-N}(\De)$
where
\be   \tilde K_{-N}(\De) =  e^{\z_{-N} V(\De)} - \z_{-N}V(\De) -1  \ee
The bound    $ \|\tilde  K_{-N}(\De) \|_{1,h_{-N}}   \leq
|\z_{-N}|^{2-\ep}  $
now follows from Lemma \ref{vbd} in the appendix.
Also  for  $|X| \geq 2 $ we have   $\| \tilde K_{-N}(X) \|_{1,h_{-N}} =
\|  K_{-N}(X) \|_{1,h_{-N}} \le  \one (| \z_{-N}|^{1- \ep} )^{|X|} $.  From
these two bounds we can deduce that for $j=-N$
\begin{equation}
\|\tilde   K_{-N}(X) \|_{G(\k),h_{-N}, \G} =  | \z_{-N}|^{2- 2\ep}
\end{equation}
Thus the theorem is established for  $j=-N$.

Next we specify the extractions  $F(K)$ from a  polymer 
activity  $K$ in the general step.  Again the extraction
is from the neutral part on small sets, but  now we only need 
$\dim(\bar K-F(K)) \ge 2$. Thus we extract only the constant
\be  ( F(K))(X) =\al(X) |X| = \bar  K(X,0)\ 1_\cS(X)
\ee

Now we suppose the theorem is true for $j$ and prove it for   $j+1$.
Starting with
 $\cE xp ( \Box  + K_j)(\La_{|j|},\f)$ we  do
 a fluctuation integral with the
measure   $\mu_{\beta C_j}$   where
\be
C_j(x-y)  =  v_0^{|j|} (0, x-y) -  v_0^{|j+1|} ( 0, (x-y)/L).
\ee
This produces   new polymer activities
$K_j^\#  = \cF(K_j)$.  Then we extract  $F_j(X)   =
\al_j(X)|X| = \bar  K_j^\#(X,0) 1_{\cS}$ as  above.
Finally we scale down to the volume $\La_{|j+1|}$.   Thus we have as in Theorem
\ref{rtheorem}
\bea && (  \mu_{\beta C} *\cE xp(\Box + K_j)(\La_{|j|}))( \f_L) \nn \\
& =& \exp \left( \sum_{X \subset \La_{|j|}} F_j(X)  \right)
\cE xp(\Box + K_{j+1})(\La_{|j+1|}, \f ) \nn \\
&=&  \exp \left( \de E_j |\La_{|j|}| \right)
\cE xp(\Box + K_{j+1})(\La_{|j+1|}, \f )
 \label{what2} \eea
where
\bea \de E_j  &=&
\sum_{X\supset \De}\alpha_j(X) \nn \\
K_{j+1} & = & \cR(K_j)  =  \cS( \cE (K_j^\# ,F(K^\#_j)) ) 
\eea
The partition function  is obtained from
(\ref{what2}) by  multiplying by  $e^{\cE_j}$ and integrating with respect
to  $ \mu_{\beta v^{|j+1|}_0}$ and has the required form
\be Z =e^{\cE_{j+1}} \int
 \cE xp ( \Box + K_{j+1})(\La_{|j+1|},\phi) d \mu_{\beta v^{|j+1|}_0}(\phi)
\ee
where
\be \cE_{j+1}=\cE_{j}+\de E_j|\La_{|j|}|.\ee

Next we  check the hypotheses of Theorem  \ref{rtheorem}
with  $\de h_j = h_j - h_{j+1}$ and $\de \k=0$.  It is easier than
before since only constants are extracted. 
 A degenerate version of Lemma
\ref{fbound} with $\al^{(2)} =0$ implies that  $F_j$ is stable
for  $(G_{\ell}(\k), h_{j+1}, f(X) ) $ and $(1, h_{j+1},\de  f(X) )$
 with $f(X)=\de f(X)=\one |\al_j(X)|$ .
Since  $|\al_j|_{\G_{-3}}\le \one
\|K_j\|_j$ is certainly small enough, the stability assumption of Theorem
\ref{rtheorem} holds.    The other  assumptions are easily checked and we 
conclude
\be  \|K_{j+1}  \|_{j+1}  \leq \one L^2  \|K_j\|_j  \ee

The leading behaviour of the RG is given by noting that
  \begin{equation}
\cR_1 (\z_j V)
=  \z_{j+1}  V   \label{rv}  \end{equation}
Indeed simple  computations give
$\cF_1 (V)  = e^{-\beta C(0)/2} V$ and  $\cE_1 (V,F(V)) = V -F(V) =V$ and
$\cS_1 V = L^2 V$. Thus  $\cR_1 ( \z_jV ) = L^2 e^{-\beta C(0)/2}\z_j V $ and
since    $  L^2 e^{-\beta C(0)/2}\z_j =\z_{j+1}   $ the claim is verified.
Because of this  we now have:
 \be \tilde  K_{j+1}=
 \cR_1( \tilde  K_j )+\cR_{\ge 2}(K_j)
\ee
If we expand   $\tilde K_j =  \sum_q k_q  $ on small sets as before
this can be written
 \be \tilde  K_{j+1}= \cR_{\ge 2}(K_j) +
  \cR_1(\tilde  K_j 1_{ \bar\cS})
+ \cR_1(\sum_{q\ne 0} k_q1_\cS)  +  \cR_1(k_01_\cS)
\label{foursum}
\ee

To show   $\|\tilde  K_{j+1} \|_{j+1}  \leq |\z_{j+1} |^{2-4\ep}$     
 we show that each of
the four terms on the right of (\ref{foursum}) can be bounded by
$ | \z_{j+1}|^{2-4\ep}/4$.  \begin{enumerate}
\item
 One checks that Theorem \ref{rtheorem} holds for $sK_j,sF_j$ with
$s\le  |\z_j|^{-1 +2\ep}$. Then by Lemma \ref{higherorderlemma}
with
$D=  |\z_j|^{-1 +2\ep} $  we
have  
\be
\|\cR_{\ge 2}(K_j)\|_{j+1}\le  \one L^2  | \z_{j}|^{1 -2\ep}\|K_j\|_j
\le  | \z_{j}|^{2-4\ep}/4  \ee 
\item  No extractions are taken from large sets so $\cR_1 (\tilde  K_j
1_{\bar\cS})=\cS_1\cF_1 (\tilde  K_j
1_{\bar\cS})$. Therefore we can use
Lemma \ref{largeset} and find
\be
\|\cR_1 (\tilde  K_j 1_{\bar\cS})\|_{j+1}\le  \one L^{-2} \|\tilde  K_j
1_{\bar\cS}\|_{j}
\leq  | \z_{j}|^{2-4\ep}/4  \leq   | \z_{j+1}|^{2-4\ep}/4 
\ee
\item
 The third term is bounded using Lemmas
\ref{better},\ref{best} just as  in the infrared section, and    we gain a
factor 
   $e^{- \beta C(0)/2 +2 N_{\beta C}} = \one L^{-\beta/ 4 \pi} $. We have 
\bea \|\cR_1 \left(\sum_{q\ne 0} k_q  \right) 1_\cS\|_{j+1}
&\le& \one L^2 \sum_{q\ne 0}
\biggl(e^{-|q|(\beta C(0)-2N_{\beta C})+\beta C(0)/2}\biggr)\
\|  \tilde K_j\|_j  \nn\\
 &\leq &    \one L^{2-\beta/4 \pi}  |\z_j|^{2- 4\ep}  \nn \\
&\leq &    \one L^{(2-\beta/4 \pi)(1-(2-4\ep))}  |\z_{j+1}|^{2- 4\ep}  \nn \\
&\leq & |\z_{j+1}|^{2-4\ep}/4   \eea
 Here we have used 
$|\z_j| \leq  \one L^{-(2- \beta/4\pi)} | \z_{j+1} |  $. 

\item  Letting $K^\dagger= \cF_1(k_0 1_{\cS}) $ we have
  $\cR_1 (k_0 1_{\cS})  = \cS_1 ( K^{\dagger}  - F( K^{\dagger} )) $.
The extraction $F$ is now defined so that
$\dim( K^{\dagger}  - F(  K^{\dagger} ) ) \geq 2$ and Lemma
 \ref{good} gives
\be
\|\cR_1(k_01_\cS)\|_{j+1}
\leq    \one
\|  K^{\dagger}  - F( K^{\dagger}) \|_{ G_{\ell}, h_{j+1}, \G_{-3}}
\ee
This is bounded by  $\one\
\|K^\dagger\|_{G_{\ell}, h_{j+1}, \G_{-3}}\le \one\|\tilde K_j\|_j$ and thus
\be \|\cR_1(k_0 1_\cS)\|_{j+1}\le
 \one  |\z_j|^{2- \ep}
\leq  | \z_{j+1}|^{2-\ep}/4  \ee
\end{enumerate}

Now  the bound on  $\|\tilde K_{j+1}\|_{j+1}$ is complete.  Next 
we need the   bound on $\de E_j $.  We have as before
\be |\de E_j| \leq  \one  \|K_j^{\#} \|_{G_{\ell}, h_{j+1}, \G_{-3}}
\leq \one \| K_j\|_j \leq  \one |\z_j|^{1-\ep}
  \ee
But we are claiming more, namely that the bound is   
actually   $ \one |\z_j|^{2-4\ep}$.  To see the improvement 
note that $\de E_j$ depends on  $ \bar K_j^\#$ where  $K_j^\#
= \cF(K_j ) = \cF_1(K_j) + \cF_{\geq 2}(K_j)$   Since 
$  \overline{ \cF_1(K_j)} =  \cF_1( \bar K_j) $
and since  $\bar V =0$ this term only depends of  $\tilde K_j$.
Thus  both terms are   $\cO( | \z_j|^{2-4\ep} )$.  We omit the  details of this
estimate. 

The analyticity of  $K_j(X, \phi) $ in $\z$ follows by observing that 
$K_{-N}(X, \phi)$ is analytic for complex  $| \z| \leq \bar \z$
and  that each  RG transformation preserves this property.
 This completes the proof of the ultraviolet theorem.

\QED
\bigskip

\appendix
\section{Appendix}

\subsection{Estimates on potentials}

\blem
 Let   $V(\De,\phi) = \int_{\De} \cos(\phi(x)) dx$ for a unit block
$\De$.  Then for any complex $\z$.
\bea  \| V(\De)  \|_{G=1,h}  & \leq&    e^{h}  \nn \\
 \| e^{\z V(\De)  }\|_{G=1,h}   &\leq & 2  \exp (|\z|e^{2h})  \eea
Furthermore for  $ 0<\ep <1$ and $|\z|$ sufficiently small
(depending on $h,\ep$)
\bea   \|  e^{ \z V(\De)}  -1 \|_{G=1,h}   &\leq&  |\z|^{1-\ep} \nn \\
 \|  e^{ \z V(\De)} -\z V(\De)  -1 \|_{G=1,h}   &\leq&  |\z|^{2-\ep}
\eea
\label{vbd}
\elem

\pr   A computation shows that  $\|V_n(\De,\phi) \| \leq 1 $ and
the first bound follows.  For the second bound we
compute  the $n^{th}$ derivative and resum
as in \cite{BDH95} and find   \be \frac {(2h)^n}{n!}
\|(e^{\z V(\De)  })_n(\phi)\|
\leq   \exp  \left( \sum_{n=0}^{\infty} \frac {(2h)^n}{n!}|\z|
\|V_n(\De,\phi)\|  \right)  \ee
Again we use  $\|V_n(\De,\phi) \| \leq 1 $  and then
take  the supremum over $\phi$ to obtain
\be \frac {(2h)^n}{n!} \|(e^{\z V(\De)  })_n\|_{G=1}
\leq   \exp  \left( |\z|e^{2h}     \right)\ee
Now multiply by $2^{-n}$ and sum over $n$ to get the result.

For the third bound we write
\be   e^{ \z V(\De)}  -1  = \frac{1}{2\pi i}  \int
\frac {e^{z \z V(\De) }}{z(z-1)}   dz  \ee
where the contour is the circle  $ |z| = |\z|^{-1+\ep/2} \ge 2$.
Since   $\| e^{z \z V(\De) }\|_{1,h}  \leq \one$
for $|\z|$ small
by the second  bound we get a bound
 $\one |\z|^{1-\ep/2} \leq |\z|^{1-\ep}$.  The fourth bound is
similar.  This completes the  proof.

\QED\bigskip

 The next  lemma is  useful in verifying
 the stability hypothesis.  Fix a unit square $\De$ and  consider a
family of quadratic
 polynomials $F(X, \De)$ defined for small sets    $X \supset  \De$
which have the form
\begin{equation}
F(X, \De)  = \al^{(0)}(X)
 + \sum_{1 \leq |a|,|b| \leq  r}\al^{(2)}_{ab}(X)  \int_{\De}  \pa^a\phi(x)
\pa^b \phi(x) dx
\end {equation}
where  $a,b$ are multi-indices.
(We could as well include a term linear in  $\pa \phi$.)
We also define
\begin{equation}
 \|  \al (X) \|_a  =  |\al^{(0)} (X) |  + a^2  |\al^{(2)}(X)|
\equiv | \al^{(0)} (X)|   + a^2  \ \sum_{ab}|\al^{(2)}_{ab}(X)|
\end{equation}

\blem Let   $\al(X)$ be supported on small sets and let
 $a=\max\{\k^{-1/2}, h\} $ for $\k\leq 1$ and $h \geq 1$.
Also let $k=\cO(1)$ be the number of small sets containing a unit
block $\De$. Then
 for all complex
$z(X)$ satisfying
\be 40 k|z(X) |  \|  \al (X) \|_a \leq 1 \ee
we have
\begin{equation}
        \|\exp  \left ( \sum _{X \supset \Delta} z(X) F(X,\Delta)
        \right ) \|_{G'(\k),h} \le  2
\end{equation}
Thus  $F$ is stable for  $ (G'(\k) , h,40k\|\al (X)\|_a ) $,
 \label{fbound}
\elem

\re  Similarly   $F$ is stable for  $ (G_{\ell}'(\k) , h, \one\|\al (X)\|_a ) $
with a larger constant  $\one$.
\bigskip

\pr  We have as above
\be \label{long} \frac {(3h)^n}{n!}  \|  \left
( \exp   ( \sum _{X \supset \Delta} z(X) F(X,\De)  )\right)_n
(\phi) \|  \leq
\exp \left (  \sum _{X \supset \Delta} |z(X)| \sum_{n=0}^{2} \frac
{(3h)^n}{n!}
\|F_n(X,\De, \phi)\|  \ \right)
\ee
Now compute the derivatives and estimate them by
\bea \label{Fbound} | F_0(X, \De, \phi)| & \leq &  |\al^{(0)}(X) |  +
 | \al^{(2)}(X) |  \| \pa \phi \|_{s,\De}^2 \nn \\
 \| F_1(X, \De, \phi) \| & \leq &
 2 | \al^{(2)}(X) |  \| \pa \phi \|_{s,\De} \nn \\
\| F_2(X, \De, \phi)\| & \leq &
2 | \al^{(2)}(X) |    \eea
Then estimate
\bea
\sum_{n=0}^{2} \frac
{(3h)^n}{n!}\| F_n(X,\De,\phi)\|
 &\le&   |\al^{(0)}(X) | + \left(\| \pa \phi \|_{s,\De}^2 +6h\| \pa \phi
\|_{s,\De} + 9h^2  \right)  | \al^{(2)}(X) |\nn\\
&\le&   |\al^{(0)}(X) | + \left(10\| \pa \phi \|_{s,\De}^2+10 h^2
\right)  |
\al^{(2)}(X) |\nn\\
&\le&   |\al^{(0)}(X) | + 10 a^2 (1 + \k \| \pa \phi
\|_{s,\De}^2 )   | \al^{(2)}(X) |\nn\\
 &\le&  40  (1/4 + \k \| \pa \phi
\|_{s,\De}^2 )   \| \al(X) \|_a
\eea
Now  since    $ 40 |z(X) |  \|  \al (X) \|_a \leq  k^{-1}$
we find
\be
\sum_{X  \supset \De} |z(X) |\sum_{n=0}^{2} \frac
{(3h)^n}{n!}\| F_n(X,
\De,\phi)\|  \leq 1/4+  \k   \| \pa \phi \|_{\De,s}^2
\ee
Using this in   (\ref{long}) yields
\be
\frac {(3h)^n}{n!}  \|  \left
( \exp   ( \sum _{X \supset \Delta} z(X) F(X,\De)  )\right)_n
 \|_{G'(\k)}   \le e^{\frac 14}
\ee
Now  multiply by $3^{-n}$  and sum over $n$
to obtain the result.

\QED

\subsection{Estimates on covariances}

Let $C_{\infty}(\si, x) $ be the covariance on $\bR^d, \ d \geq 2$ defined by
\be C_{\infty} (\si , x) = (2\pi)^{-d}\int_{\bR^d}\ dp\
\frac{ e^{ipx}}
{p^2}\ [(e^{p^4}+\si)^{-1}-(e^{L^4p^4}+\si)^{-1}]
\ee

\blem\label{cbound}
\begin{enumerate}
\item There is  $\sigma_0 = \cO(1)$ such that for $|\sigma| \leq
\sigma_0$ and  any multi-index $\beta$  there are constants
$c_1,c_2$  such that
\bea  \label{c1c2c3}
|\pa^\beta C_{\infty} (\si ,x)| &\le & c_1 \exp ( - |x|/L) \nn \\
\int |\pa^\beta C_{\infty}( \si ,x)|dx &\le & c_2
\eea
The constant $c_1  =   \one \log L$  for $d=2, \beta = 0$, but  may be chosen
independent of
$L$ otherwise.  We also have  $c_2 \leq  \one \int_1^L s^{1- |\beta|} ds$.
\item In $d=2$,
\be  C_{\infty}(\si,0)=  \frac{ \log L}{ 2\pi(1+\si)}  \ee
\end{enumerate} \elem

\pr We rewrite the covariance and its derivatives as
\bea
\pa^{\beta} C_{\infty}(\si, x)
 &=& (2\pi)^{-d}\int^{L}_1 \ ds\
\int_{\bR^d}\ dp\
\frac{e^{ipx}}{p^2}(ip)^\beta\ (- \frac{\pa} {\pa s})
 (e^{s^4p^4}+\si)^{-1}\nn\\
 &=& (2\pi)^{-d}\int^{L}_1 \ ds\
\int_{\bR^d}\ dp\
\frac{ e^{ipx}}{p^2}(ip)^\beta
 \frac{(4s^3 p^4e^{s^4p^4})}{(e^{s^4p^4}+\si)^{2}}\nn \\
&=&  4(2\pi)^{-d} \int^{L}_1 \frac { ds}{  s^{d-1+|\beta|}}\ \int_{\bR^d}\
dp\ e^{is^{-1}px}\left[ (ip)^\beta p^2 \
 \frac{e^{p^4}}{(e^{p^4}+\si)^{2}}  \right]
\label{sintegral}
\eea
The  function in brackets  is analytic, bounded and integrable in
the strip $|\imag (p)| \leq1 $ around the real axis when $|\si|$ is small.
Therefore we can   shift  the
$p$ integral one unit in an imaginary direction and
exhibit the exponential decay in
$x$.  We find
\be
|\pa^\beta C_{\infty} (\si,x)| \le \one  \int^{L}_1 \ \frac{ds}
{s^{d-1+|\beta|}}\ e^{-s^{-1}|x|}
\ee
and the bounds (\ref{c1c2c3}) follow.
In $d=2$ we compute
\bea
C_{\infty}( \si,0)  &=& \pi^{-2} \int^{L}_1 \frac{ ds}{s}\
\int^\infty_0 \
2\pi rdr\ \frac{r^2e^{r^4}}{(e^{r^4}+\si)^{2}}\\
 &=& \frac{\log L}{2\pi(1+\si)}
\eea
This completes the proof.

\QED
\bigskip

Now let  $C^M(\si, x)$ be the covariance on  $\La_M$ as defined in
(\ref{Mcovariance})

\blem\label{covariancebound}   Let    $|\si|\le \sigma_0$ .
\begin{enumerate}
\item For any multi-index $\beta$ and  $|x| \leq L^M/2$
\bea  \label{newc1c2c3}
|\pa^\beta C^M (\si ,x)| &\le & \one  c_1 \exp ( -  |x|/L) \nn \\
\int |\pa^\beta C^M( \si ,x)|dx &\le &  \one  c_2
\eea
\item In $d=2$,
\be  C^M(\si,0)=  \frac{ \log L}{ 2\pi(1+\si)} +   \one e^{-L^{M-1}/2}\ee
\end{enumerate} \elem

\pr   We have the representation
\be  C^M(\si, x)  =  \sum_{n \in \bZ^2}   C_{\infty}(\si, x+ n L^M)  \ee
This follows since both sides are doubly periodic with period
$L^M$, and they have the same Fourier coefficients, namely
$p^{-2} ((e^{p^4}+\si)^{-1} - (e^{L^4p^4}+\si)^{-1} ) $ for $p \neq 0$ and
$0$ for $p=0$.
The terms in the sum are estimated by the previous lemma
and we obtain all the stated  results.

\QED
\bigskip

\nind{\bf  Acknowledgement:}  We thank David Brydges for  his
encouragement and many helpful conversations.

\end{document}